\documentclass[preprint,prd,amsmath,amssymb,nofootinbib]{revtex4}
\usepackage{changes}
\usepackage{amsfonts}
\usepackage{amsmath,graphicx,color}
\usepackage{epstopdf}
\usepackage{threeparttable}

\newcommand{\be}{\begin{equation}}
\newcommand{\ee}{\end{equation}}
\newcommand{\bea}{\begin{eqnarray}}
\newcommand{\eea}{\end{eqnarray}}
\newcommand{\ba}{\begin{array}}
\newcommand{\ea}{\end{array}}
\newcommand{\bi}{\begin{itemize}}
\newcommand{\ei}{\end{itemize}}

\newcommand{\mcn}{{\mathcal N}}

\newcommand{\nslash}{\kern 0.2 em n\kern -0.50em /}
\newcommand{\kslash}{\kern 0.2 em k\kern -0.45em /}
\newcommand{\qslash}{\kern 0.2 em q\kern -0.45em /}
\newcommand{\pslash}{\kern 0.2 em p\kern -0.50em /}
\newcommand{\rslash}{\kern 0.2 em r\kern -0.50em /}
\newcommand{\sslash}{\kern 0.2 em s\kern -0.50em /}
\newcommand{\Sslash}{\kern 0.2 em S\kern -0.50em /}
\newcommand{\Pslash}{\kern 0.2 em P\kern -0.50em /}
\newcommand{\Dslash}{\kern 0.2 em D\kern -0.65em /\kern 0.15em}

\begin{document}
\title{ Photoproduction of hidden-bottom pentaquark and related topics}

\author{Xu Cao$^{1,2}$\footnote{caoxu@impcas.ac.cn},
Feng-Kun Guo$^{3,2}$\footnote{fkguo@itp.ac.cn},
Yu-Tie Liang$^{1,2}$\footnote{liangyt@impcas.ac.cn},
Jia-Jun Wu$^{4}$\footnote{wujiajun@ucas.ac.cn},
Ju-Jun Xie$^{1,2}$\footnote{xiejujun@impcas.ac.cn},
Ya-Ping Xie$^{1,2}$\footnote{xieyaping@impcas.ac.cn},
Zhi Yang$^{1,2}$\footnote{Corresponding author: zhiyang@impcas.ac.cn},
Bing-Song Zou$^{3,2}$\footnote{zoubs@itp.ac.cn}}
\affiliation{$^1$Institute of Modern Physics, Chinese Academy of Sciences,
Lanzhou 730000, China}
\affiliation{$^2$University of Chinese Academy of Sciences, Beijing 100049,
China}
\affiliation{$^3$CAS Key Laboratory of Theoretical Physics, Institute of
Theoretical Physics, Chinese Academy of Sciences, Beijing 100190, China}
\affiliation{$^4$School of Physical Sciences, University of Chinese Academy of
Sciences, Beijing 100049, China}
\begin{abstract}
  Due to the discovery of the hidden-charm pentaquark $P_c$ states by the LHCb collaboration, the interests on the candidates of hidden-bottom pentaquark $P_b$ states are increasing. They are anticipated to exist as the analogues of the $P_c$ states in the bottom sector and predicted by many models. We give an exploration of searching for a typical $P_b$ in the $\gamma p \to \Upsilon p$ reaction, which shows a promising potential to observe it at an electron-ion collider. The possibility of searching for $P_b$ in open-bottom channels are also briefly discussed. Meanwhile, the $t$-channel non-resonant contribution, which in fact covers several interesting topics at low energies, is systematically investigated.
\end{abstract}
\maketitle

%%%%%%%%%%%%%%%%%%%%%%%%%%%%%%%%%%%%%%%%%%%%%%%%%%%%%%%%%%%%%%%%%%%%%%%%%%%%%%%%%%%%%%%%%%%%%%%%%%%%%%%%%%%%%%%%%%%%%%%
\section{Introduction} \label{sec:intro}

Since the discovery of the $X$(3872) by the Belle collaboration~\cite{Choi:2003ue}, a rich spectrum of exotic states has been emerging, see comprehensive reviews for references~\cite{Chen:2016qju,Guo:2017jvc,Lebed:2016hpi,Esposito:2016noz,Olsen:2017bmm,Liu:2019zoy,Brambilla:2019esw,Guo:2019twa}. They not only shed new insights into the study of hadron spectrum and structure, but also deepen our understanding of nonperturbative properties of Quantum Chromodynamics (QCD). Among these states, the charged $Z_c$(3900) and $Z_c$(4020) found respectively in the $J/\psi \pi^{\pm}$~\cite{Ablikim:2013mio,Liu:2013dau} and $h_c \pi^{\pm}$~\cite{Ablikim:2013wzq} systems
seem to be surely exotic since they must contain at least one additional light quark and anti-quark pair besides the hidden pair of $c \bar{c}$ to match the electric charge.
Their partners in the bottomonium sector, namely the $Z_b$(10610) and $Z_b$(10650), were firmly established by Belle in several different decay modes~\cite{Belle:2011aa}. The spin and parity of these states are determined unambiguously to be $1^+$ by the amplitude analysis of BESIII~\cite{Collaboration:2017njt} and Belle~\cite{Garmash:2014dhx}, except for the $Z_c$(4020), which is believed to be of the same quantum numbers by most of the models. Their masses are very close to the $S$-wave thresholds of the corresponding open-flavor channels $D \bar{D}^{(*)}$ and $B \bar{B}^{(*)}$, respectively.
As for their strange partner $Z_s$, so far the BESIII Collaboration did not find a signal in the $\phi \pi$ spectrum of $e^+e^-\rightarrow\phi\pi\pi$~\cite{Ablikim:2018ofc}.

In the baryon sector, the hidden-strangeness pentaquark $P_s$ states containing only light quarks are expected in constituent quark models~\cite{Huang:2018ehi,Liu:2018nse}, and in  models considering the QCD van der Waals force~\cite{Gao:2000az,He:2018plt}. But they are not explicitly found at present after a long searching for them in $\pi N$ and $\gamma N$ reactions~\cite{Cao:2017njq}. Other reactions and decays were suggested to study them from the theoretical side~\cite{Cao:2014vma,Cao:2018vmv,Cao:2014mea,Lebed:2015dca,He:2017aps,An:2018vmk,Gao:2017hya}. Interestingly, no narrow peaks were found in total cross section of near threshold $\gamma p\to \phi p$, but a non-monotonic structure, found in the differential cross section by LEPS Collaboration~\cite{Mibe:2005er}, would imply a very wide $\sim$ 500 MeV states~\cite{Kiswandhi:2010ub,Kiswandhi:2011cq}.
There is also no any evident signal in the $\phi p$ energy spectrum of the process $\Lambda_c^+\to \phi p \pi^0$~\cite{Pal:2017ypp}, which was shown in Ref.~\cite{Xie:2017mbe} to be not a good choice for the search of $P_s$ due to the presence of triangle singularities (for a recent review, see Ref.~\cite{Guo:2019twa}) and the tiny phase space.
However, in the charm sector, the astonishing observation of $P_c$ states by the LHCb Collaboration~\cite{Aaij:2015tga,Aaij:2019vzc} has provided us an insightful place to study the exotic baryons in the charm sector, the existence of which were anticipated by several models~\cite{Wu:2010jy,Wu:2010vk,Wang:2011rga,Yang:2011wz}.
The photoproduction reactions of these states with two-body final states, first proposed in Ref.~\cite{Wang:2015jsa} and followed by other works~\cite{Karliner:2015voa,Kubarovsky:2015aaa}, are an exceptional platform to exclude their non-resonant possibility, because the on-shell conditions required by the triangle singularities discussed in Refs.~\cite{Liu:2015fea,Guo:2015umn,Guo:2016bkl,Bayar:2016ftu,Liu:2019dqc} cannot be satisfied.
The upper limit of the $P_c$ photoproduction cross section in $\gamma p \to J/\psi p$ was determined recently by the GlueX Collaboration~\cite{Ali:2019lzf}, constraining the branching ratios of the $P_c$ decays into the $J/\psi p$ mode together with the results at LHCb~\cite{Cao:2019kst}.
Due to the null results in the GlueX data, double polarization observables were proposed to be a benchmark in the search of pentaquark photoproduction~\cite{Winney:2019edt}. Although the nature of these exotic states is under discussion~\cite{Chen:2019asm,Guo:2019fdo,Guo:2019kdc,Eides:2019tgv,Wang:2019got,Ali:2019clg,Burns:2019iih}, they motivated the speculation from heavy-quark spin symmetry that there should be seven molecular pentaquarks in two spin multiplets~\cite{Liu:2019tjn,Xiao:2019aya,Du:2019pij}. Motivated by the heavy quark flavor symmetry for the potential between heavy mesons and baryons, the correspondence of these states in the bottom sector, label as $P_b$ here, are expected to be surely existing~\cite{Wu:2010rv,Xiao:2013jla,Karliner:2015voa,Karliner:2015ina}. Unlike the $P_c$, they cannot be produced through the decay of heavier baryons. Therefore, they can only be directly produced in high-energy processes, such as the $ep$,  $\gamma p$ scattering and the $pp$ collisions.

In this paper we will discuss the possibilities of searching for one of typical $P_b$ states, the bottom analogs of $P_c$, in the photoproduction of the the bottomonium channel $\gamma p \to \Upsilon p$ at electron-ion colliders (EICs). To this end, we first explore the non-resonant contribution to the $\gamma^* p \to \Upsilon p$ in Sec.~\ref{sec:Pomeron}. This is very meaningful on its own right because several subjects are relevant to it. The detailed investigation of the $P_b$ contribution is presented in Sec.~\ref{sec:Pb}. At last we finish with a short summary in Sec.~\ref{sec:summ}.

%%%%%%%%%%%%%%%%%%%%%%%%%%%%%%%%%%%%%%%%%%%%%%%%%%%%%%%%%%%%%%%%%%%%%%%%%%%%%%%%%%%%%%%%%%%%%%%%%%%%%%%%%%%%%%%%%%%%%%%
\section{Non-resonant contribution} \label{sec:Pomeron}

\begin{figure}
\centering
\includegraphics[width=0.35\textwidth,clip]{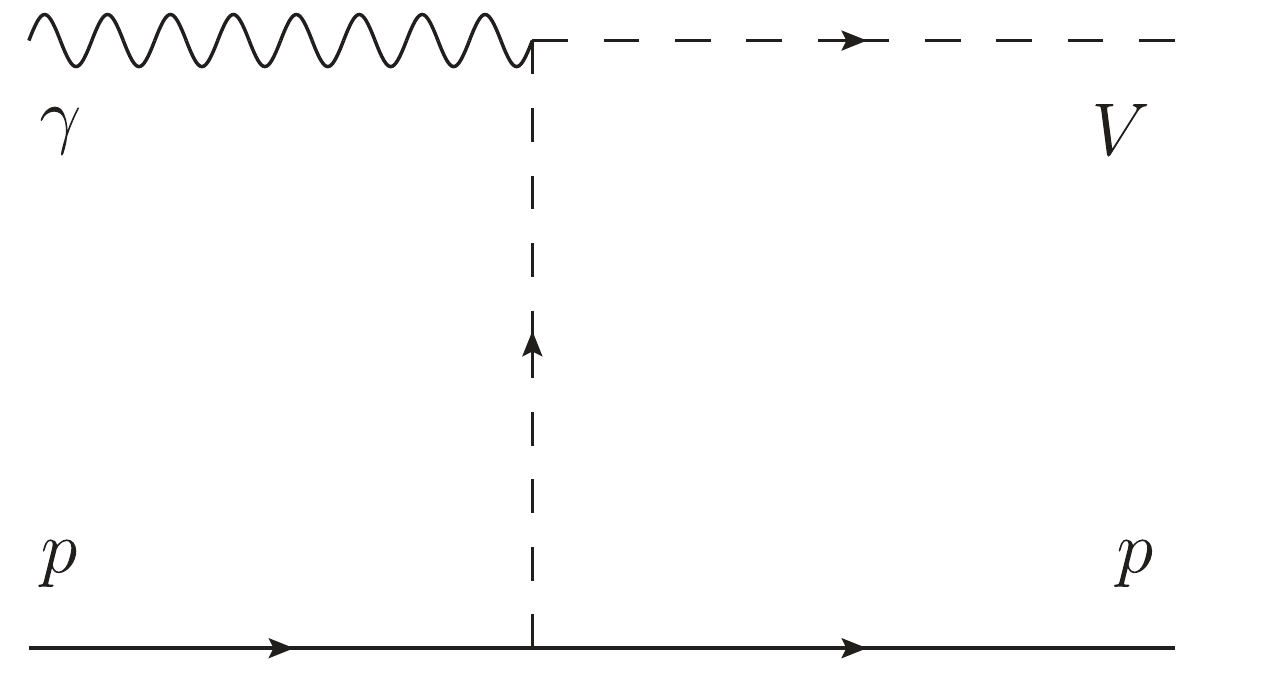}\qquad
\includegraphics[width=0.35\textwidth,clip]{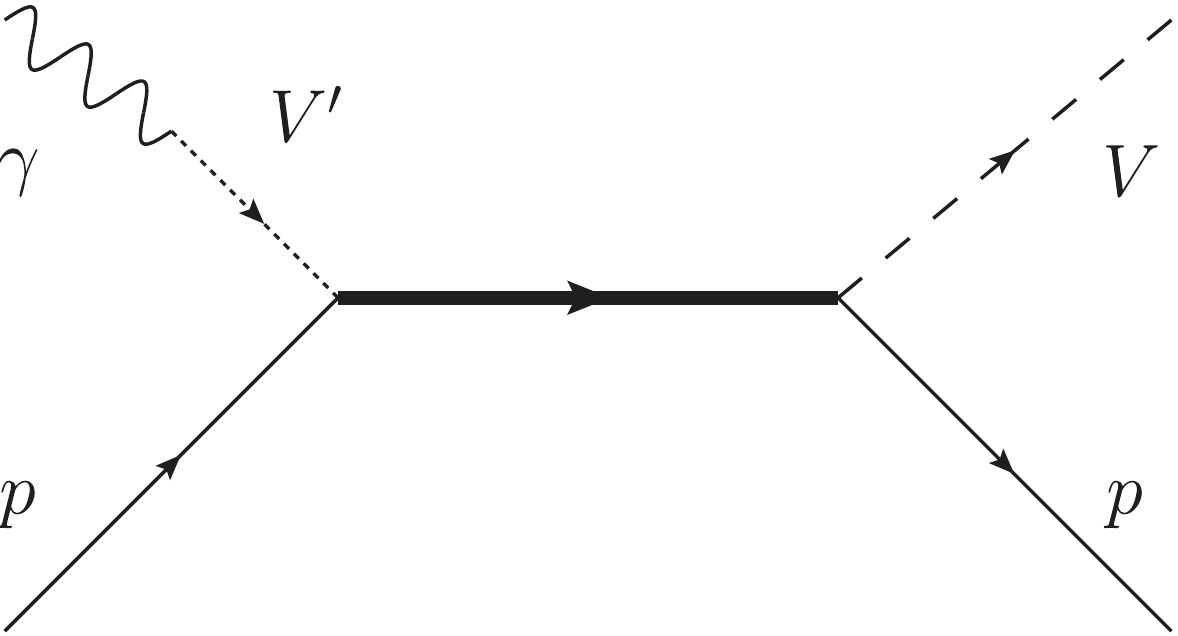}
\caption{Feynman diagrams for $\gamma p\to \Upsilon p$. Left: the $t$-channel contribution with the Pomeron or two-gluon exchange. Right: the $P_b$ production in the $s$-channel. $V$ labels the $\Upsilon$ meson. $V'$ denotes the possible vector meson, including $\rho$, $\omega$, $\phi$, and $\Upsilon$, in the vector-meson dominance model.}
\label{fig:feynman}
\end{figure}
\begin{figure}
  \begin{center}
  {\includegraphics*[width=0.6\textwidth]{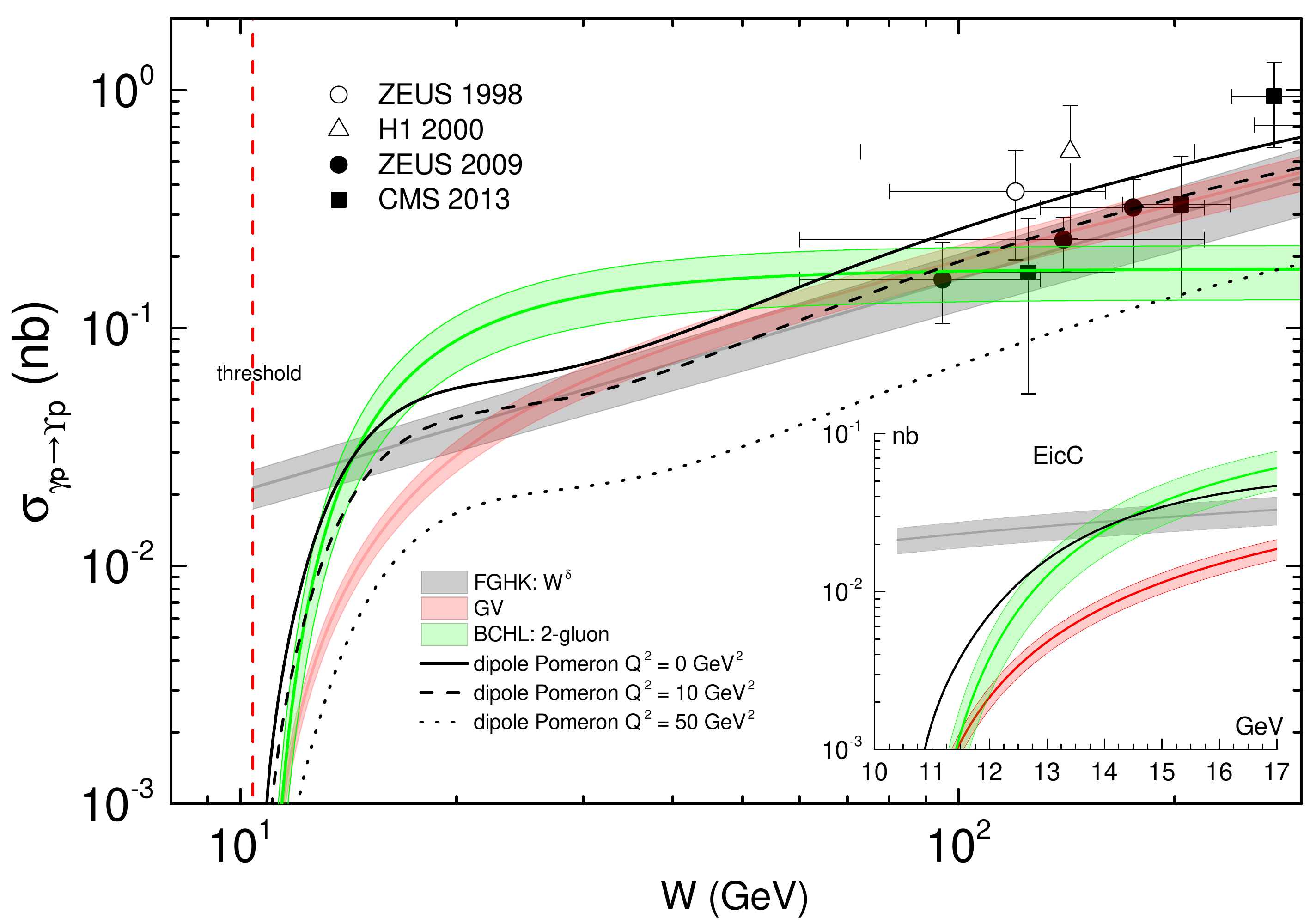}}
    \caption{The non-resonant contribution to the $\gamma^* p \to \Upsilon p$ as a function of the c.m. energy of $\gamma^* p$, $W$. The data are from ZEUS~\cite{Breitweg:1998ki,Chekanov:2009zz}, H1~\cite{Adloff:2000vm} and CMS~\cite{CMS:2016nct}. The data above 300 GeV from CMS~\cite{CMS:2016nct} and LHCb~\cite{Aaij:2015kea} were used in the fit in order to determine the overall normalization but are not shown. The models include the DVMP empirical formula (FGHK)~\cite{Favart:2015umi}, two-gluon exchange model (BCHL)~\cite{Brodsky:2000zc}, the parameterization in Ref.~\cite{Gryniuk:2016mpk} (GV), and the soft dipole Pomeron~\cite{Martynov:2001tn,Martynov:2002ez}. The cross sections of $\gamma^* p \to \Upsilon p$ under $Q^2 = 10$ (dashed line) and 50 GeV$^2$ (dotted line) are also given by the soft dipole Pomeron model. The inserted subplot on right bottom enlarges the energy region covered by the proposed EicC.
    \label{fig:Upsilon_all}}
  \end{center}
\end{figure}

The main purpose of studying the photo- and electroproduction of vector heavy quarkonia off the nucleon is to study the gluon component within the nucleon probed by heavy quarks. The low energies are also important for several other topics which are critically relevant. First, the near-threshold region would provide clue for the quarkonium-nucleon interaction. The measured cross sections have been used to extract the $J/\psi p$ scattering length~\cite{Gryniuk:2016mpk,Strakovsky:2019bev}, whereas the $\Upsilon p$ scattering length is rarely known due to the lack of data. Second, it was proposed to be a promising platform to probe the trace anomaly term in the QCD energy-momentum tensor and the proton mass decomposition, resulting into a deep exploration of the origin of the nucleon mass~\cite{Kharzeev:1995ij,Hatta:2018ina}.

Since the discovery of the $J/\psi$, its photoproduction has attracted plenty of interests both from experimental and theoretical aspects. Because the bottom quark is heavier than the charm one, the $\Upsilon$ photoproduction has its own merits. The multipole expansion~\cite{Kharzeev:1995ij,Fujii:1998tk} converges more quickly. Relative uncertainties of the current quark mass and the running coupling constant are much smaller. This is an essential advantage for the theoretical calculation because the amplitudes are expected to be proportional to powers of these quantities. At last but not at least, it is ideal to search for hidden-bottom pentaquark candidates. The GlueX Collaboration has searched for the $P_c$ in the near-threshold region of the $\gamma p \to J/\psi p$~\cite{Ali:2019lzf} as mentioned above. The $P_b$ states, whose lowest mass in many theoretical models is expected to be lying above $\Upsilon p$ production threshold, making $\gamma p \to \Upsilon p$ as a perfect place to hunt for them. However, the data of the $\Upsilon$ production below 100~GeV has never been measured up to now, and it becomes one of the main concerns of the proposed Electron-Ion Collider in China (EicC), as proposed in the white paper~\cite{CAO:2020EicC}.

In order to explore the possibility of studying the $\Upsilon$ production at relative low energies, we need to estimate its cross section as a premise, with the help of a reliable model to extrapolate from high to low energies. The non-resonant contribution would come from the $t$-channel two-gluon or Pomeron exchange, as shown in Fig.~\ref{fig:feynman}. A rough evaluation of total cross section reads as
\be \label{eq:sigmaDVMP}
  \sigma(\gamma^* p \to V p) = \mathcal{N} W^{\delta(Q^2)} = \mathcal{N} W^{\alpha+\beta \ln (Q^2+M_V^2)} \quad ,
\ee
which is suggested by the empirical formula of the deeply virtual meson production (DVMP) $\gamma^* p \to V p$~\cite{Favart:2015umi}. Here  $M_V$ is the $\Upsilon$ mass, $W$ the  $\gamma p$ center of mass (c.m.) energy and $Q^2$ the photon virtuality. The advantage of this simple parameterization is that it is applicable to all DVMP processes with proper $Q^2$ dependence. The parameters $\alpha$ and $\beta$ have been determined by the DVMP data to be $\alpha = 0.31 \pm 0.02$ and $\beta = 0.13 \pm 0.01$ by Favart, Guidal, Horn and Kroll (FGHK)~\cite{Favart:2015umi}. Correspondingly, $\delta(Q^2=0) = 0.89 \pm 0.05$, confronted with the perturbative QCD prediction $\delta \sim 1.7$~\cite{Frankfurt:1998yf} and ZEUS results $\delta = 0.69 \pm 0.02 \pm 0.03$~\cite{Chekanov:2002xi}.
The normalization $\mathcal{N}$ is determined by the data of $\gamma p \to \Upsilon p$ at high energies to be $2.62 \pm 0.38$, where the experimental uncertainty of $W$ is not taken into account. The result is shown as the grey band in Fig.~\ref{fig:Upsilon_all}, together with those from a few other models. As can be seen, the general trend of high energy data with large errors follows the exponential behavior in Eq.~(\ref{eq:sigmaDVMP}). Note that the data above 300~GeV up to 2~TeV from CMS~\cite{CMS:2016nct} and LHCb~\cite{Aaij:2015kea} were used in the fit in order to determine the overall normalization; they also follow the exponential behavior though not shown in the figure.

Gryniuk and Vanderhaeghen (GV) adopted the following parametrization for the cross section~\cite{Gryniuk:2016mpk}:
\be \label{eq:Gryniuk}
\sigma (\gamma p \to V p) = \left( \frac{e f_V}{M_V} \right)^2 \frac{\mathcal{N}}{2 W \, q_{\gamma p}} \left( \frac{q_{V p}}{q_{\gamma p}} \right) \, \left( 1 - \frac{\nu_{\rm el}}{\nu} \right)^{b_{\rm el}}  \left( \frac{\nu}{\nu_{\rm el}} \right)^{a_{\rm el}} \quad ,
\ee
where $f_V$ is the vector meson decay constant, $\nu = (W^2 - m_p^2 - M_V^2)/2$, and $\nu_{\rm el}= m_p M_V$. The quantities $q_{\gamma p}$ and $q_{V p}$ denote the magnitude of the three momenta in the c.m. frame of initial and final states, respectively. The parameters $a_{\rm el} = 1.27 \pm 0.17$ and $b_{\rm el} = 1.39 \pm 0.01$ are determined by the data of $\gamma p \to J/\psi p$. Good agreement is found after fitting the normalization $\mathcal{N} = 0.014 \pm 0.002$ to the high energy data, as shown by the red band in Fig.~\ref{fig:Upsilon_all}. This simple parametrization generally preserves the exponential trend at high energies, and surprisingly agrees very well with the data above 100~GeV.

The two-gluon exchange model proposed by Brodsky, Chudakov, Hoyer and Laget (BCHL) suggests the following $t$-dependent cross sections~\cite{Brodsky:2000zc}:
\be \label{eq:2gluon}
%\frac{\textrm{d} \sigma_{2g}}{\textrm{d} t} (\gamma p \to V p) = \mcn_{2g} \frac{(1-x)^2}{R^2M_v^2} \nu (W^2-m_p^2)^2 e^{b\,t}
\frac{\textrm{d} \sigma_{2g}}{\textrm{d} t} (\gamma p \to V p) = \mcn_{2g} \frac{(1-x)^2}{R^2M_V^2} \frac{1}{16 \pi} e^{b\,t} ,
\ee
% $16 \pi \nu (W^2-m_p^2)^2 = 1$
with $R = 1$ fm, $x = (2m_p M_V + M_V^2)/(W^2 - m_p^2)$, and the transfer-momentum squared  $t$. We use the slope parameter $b=1.13$ GeV$^{-2}$ in the original scheme, which is compatible with the measured one $b=1.25 \pm 0.20$ GeV$^{-2}$ at $W=11$~GeV for the $J/\psi$ production~\cite{Gittelman:1975ix}. The corresponding result is shown as the green band in Fig.~\ref{fig:Upsilon_all}. The normalization $\mcn_{2g}$ is adjusted to the data around 100~GeV because obviously this model cannot describe the data at high energies. The same authors also proposed the form of three-gluon exchange with an unknown normalization. It is premature to discuss such a contribution at present because of lack of data below 100~GeV.

Several Pomeron models have been constructed~\cite{Laget:1994ba}, but few of them have been used to study the case of the $\Upsilon$. The soft dipole Pomeron model, put forward by Martynov, Predazzi and Prokudin~\cite{Martynov:2001tn,Martynov:2002ez}, preserves unitarity bounds with a double Regge pole with an intercept equal to 1. By fitting to all the available data of $\gamma^* p \to V p$ at that moment, the model predicts the behavior of $\gamma^* p \to \Upsilon p$, which is consistent with the measured data afterwards, see the black curves in Fig.~\ref{fig:Upsilon_all}. Besides the usual exponential tendency at high energies, additional small fluctuations are observed. The shoulder shape around 20~GeV is caused by a Regge pole mainly contributing to low energies.
The trough around 30~GeV is from the interference between two Regge poles. We also show the $Q^2$ dependence of the cross sections, which tend to be more moderate with larger $Q^2$ as expected from  the $(Q^2 +M_V^2)^{-1}$ behavior. Sibirtsev \emph{et al.} also concluded that two Regge trajectories were required to describe the data of $\gamma^* p \to J/\psi p$ over a wide energy range after comparing various models~\cite{Sibirtsev:2004ca}. This is different from most of the Pomeron models with only one Regge trajectory~\cite{Wang:2015jsa,Blin:2016dlf}.

Figure~\ref{fig:Upsilon_all} shows that various models can describe the data at high energies comparably well, except the two-gluon exchange model, which is designed to focus on the near-threshold region. However, the inserted subfigure in Fig.~\ref{fig:Upsilon_all} shows that the deviations between different models are large at low energies, which are covered by the proposed EicC. The empirical formula of DVMP, as a guideline and a rough upper limit, does not take into account the influence from phase space, which is significant at low energies as one can easily anticipate. The soft dipole Pomeron model overlaps with the two-gluon exchange one within uncertainties, but is larger in the very close-to-threshold range. The GV parametrization is smaller than the other models below 20~GeV.

In a short conclusion, the soft dipole Pomeron model and the GV parametrization are both compatible with high energy data and give the expected behavior of the phase space at low energies. So they serve as a good input for the study of the non-resonant contribution to the $e p \to ep \Upsilon$ process. Because the parameters in the soft dipole Pomeron model is from a global fit to all the data, in the next section we will use it as the non-resonant contribution to $\gamma p \to \Upsilon p$. We also use the empirical formula from DVMP as a crude estimation of the upper limit of the non-resonant contribution. Besides, other models that  are available to calculate the cross section of the $\gamma^* p \to J/\psi p$ can also be extended to the case of the $\gamma^* p \to \Upsilon p$. However, most of them have more undetermined parameters owing to lack of data of the $\Upsilon$ production, so we do not consider them at present.

\section{$P_b$ as a resonance in photoproduction} \label{sec:Pb}

\begin{table}[b]
  \begin{center}
    \caption{Parameters of $P_b$ in models. Here we only list the $P_b$ with the mass around 11.12 GeV and other $P_b$ is not included.}
  \label{tab:Pb}
    \begin{threeparttable}
  %\footnotesize
 \begin{tabular}{c|c|c|c|c}
\hline\hline
       $P_b$           &Mass $M$ (GeV)               &Width $\Gamma$(MeV)       & $\Gamma(P_b \to \Upsilon p)$ & $\mathcal{B}(P_b \to \Upsilon p)$     \\
\hline
   J. J. Wu et al.~\cite{Wu:2010rv}   & $11.10$  &   1.33           & 0.51                             & 0.38 \\
   Karliner\&Rosner~\cite{Karliner:2015voa,Karliner:2015ina}  & 11.14  & 39\tnote{$\ddag$}~~or 61\tnote{$\dag$}  & --- & 0.1 \\
   Huang et al.~\cite{Huang:2015uda,Huang:2018wed} & 11.09 - 11.14\tnote{$\maltese$}    &   7.0            & 4.4           & 0.63  \\
   Lin et al.~\cite{Lin:2018kcc}      & ---      &   30-300         & ---                              & 0.0003-0.0013  \\
   Yang et al.~\cite{Yang:2018oqd}    & 11.14    &   ---            & ---                              & ---  \\
   Xiao et al.~\cite{Xiao:2013jla}    & 10.96-11.022  &   2-110     & ---                              & ---  \\
   Shen et al.~\cite{Shen:2017ayv}    & 11.120   &   25             & ---                              & ---  \\
   Gutsche et al.~\cite{Gutsche:2019mkg}    & 11.125  & ---         & 3.27 & ---  \\
   Gutsche et al.~\cite{Gutsche:2019mkg}    & 11.13   & ---         & 6.57 & ---  \\
\hline\hline
   \end{tabular}
  \begin{tablenotes}
  \item[$\maltese$] If all closed channels included, it is 10.304 ($1/2^-$) and 10.382 ($3/2^-$).
  \item[$\dag$] Roughly estimation from phase space ratio $\Gamma(P_b)/\Gamma(P_c) = k_{\rm out}(P_b)/k_{\rm out}(P_c)$.
  \item[$\ddag$] Assume the same width with $P_c$(4450) at LHCb.
  \end{tablenotes}
\end{threeparttable}
\end{center}
\end{table}

We list the properties of a typical $P_b$ predicted by various phenomenological models which used the $P_c$ as inputs in Tab.~\ref{tab:Pb}. We do not attempt to collect all the models here because of the still increasing literature. We would like to point out that nearly all models predict a resonant state with a mass around 11.12~GeV which couples to the $\Upsilon p$ channel, while the total width differs due to detailed constructions of the models, ranging from 30~MeV to 300~MeV. In this paper we will adopt the mass of 11.12~GeV with two possible width values 30~MeV and 300~MeV. Later on they are debbed as the narrow $P_b$ and the wide $P_b$, respectively. The spin $J = 1/2$ is used here as a representative choice. Other $P_b$ states with different quantum numbers can be similarly calculated since the production cross section is proportional to $2J+1$ in our prescription.

The production cross section of the exotic $P_b$ in the reaction $\gamma p \to \Upsilon p$, as shown by the Feynman diagram in Fig.~\ref{fig:feynman}, can be written as
\be
\label{eq:sigmaR}
  \sigma_{P_b} = \frac{2\,J+1}{(2\,s_1+1)(2\,s_2+1)} \frac{4\,\pi}{k_{\rm in}^2} \frac{\Gamma^2}{4} \frac{\mathcal{B}(P_b \to \gamma p) \, \mathcal{B}(P_b \to \Upsilon p)}{(W-M)^2+\Gamma^2/4}.
\ee
Here $k_{\rm in}$ is the magnitude of the initial-state three momentum in the c.m. frame, and $s_1$ and $s_2$ are the spins of initial photon and proton, respectively. Because the mass $M$ of $P_b$ is very large, this formula is a very good approximation even for the wide $P_b$. If assuming that the $P_b \to \gamma p$ is dominated by only the heavy vector meson in the vector meson dominance model, e.g., $V' = \Upsilon$ in Fig.~\ref{fig:feynman}, the branching ratio $\mathcal{B}(P_b \to \gamma p)$ is proportional to $\mathcal{B}(P_b \to \Upsilon p)$~\cite{Karliner:2015voa,Kubarovsky:2015aaa,Cao:2019kst}:
\be
\mathcal{B}(P_b \to \gamma p)= \frac{3\,\Gamma(\Upsilon \to e^+e^-)}{\alpha M_{\Upsilon}}  \frac{k_{\rm in}}{k_{\rm out}} \mathcal{B}(P_b \to \Upsilon p)  \,,
% f_L \lf( \rg)^{2L+1}
\ee
which has assumed the lowest orbital excitation $L = 0$ between the $\Upsilon$ and the proton. Here $\alpha$ is the fine structure constant, $k_{\rm out}$ is the magnitude of final-state three momentum in the c.m. frame, and the dilepton width $\Gamma(\Upsilon \to e^+e^-) = 1.34$~keV~\cite{Tanabashi:2018oca}. As a result, we have $\sigma_{P_b} \propto \mathcal{B}^2(P_b \to \Upsilon p)$. It shall be noted that the intermediate vector meson $V' = \Upsilon$ in Fig.~\ref{fig:feynman} is highly off-shell, so a form factor would be present with a possible strong suppression, as pointed out in Ref.~\cite{Wu:2019adv}. At present, the branching fraction $\mathcal{B}(P_b \to \gamma p)$ is not directly measured so the magnitude of this form factor is unknown. As a result, $\mathcal{B}(P_b \to \gamma p)$ above needs to be understood as an effective branching ratio with this factor absorbed. Recently, the measurement of GlueX at JLab Hall-D has given the upper limit of $\mathcal{B}(P_c^+\to J/\psi p)$ to be several per cent without considering this off-shell factor. The LHCb results indicate a stringent lower limit of $\mathcal{B}(P_c^+\to J/\psi p)$ to be $0.05\% \sim 0.5\%$~\cite{Cao:2019kst}. We use these values of $P_c$ as a reference and adopt $0.5\% < \mathcal{B}(P_b \to \Upsilon p) < 5\%$ for $P_b$. The calculated values in most of the models in Tab.~\ref{tab:Pb} are within this chosen range, except one of them approaching to about 0.01\%~\cite{Lin:2018kcc}.

The non-resonant contribution studied in Sec.~\ref{sec:Pomeron} is considered as the smooth background of $P_b$. The interference effect between them in the total and differential cross sections is not significant because the $t$-channel Pomeron exchange contributes only to the forward angles while the $s$-channel resonances are present in full angles. The calculation of $\gamma p \to J/\psi p$ confirm this  expectation~\cite{Wu:2019adv}. The hereafter error bands are from the uncertainty of non-resonant contribution but does not include the errors of the mass $M$ and width $\Gamma$ of $P_b$, just because it is too premature to consider them at this stage.

\begin{figure}
  \begin{center}
  {\includegraphics*[width=0.47\textwidth]{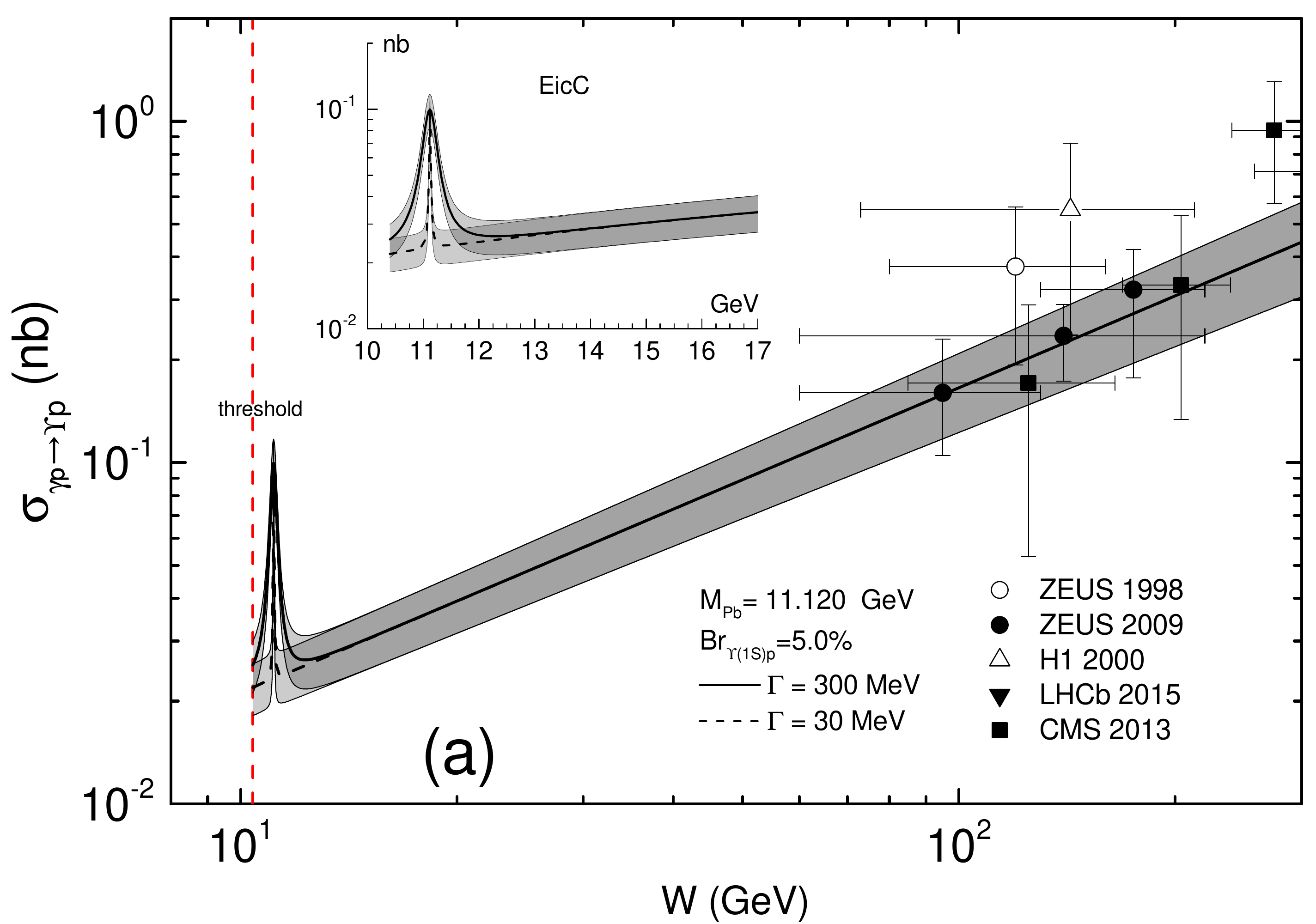}}\quad
  {\includegraphics*[width=0.47\textwidth]{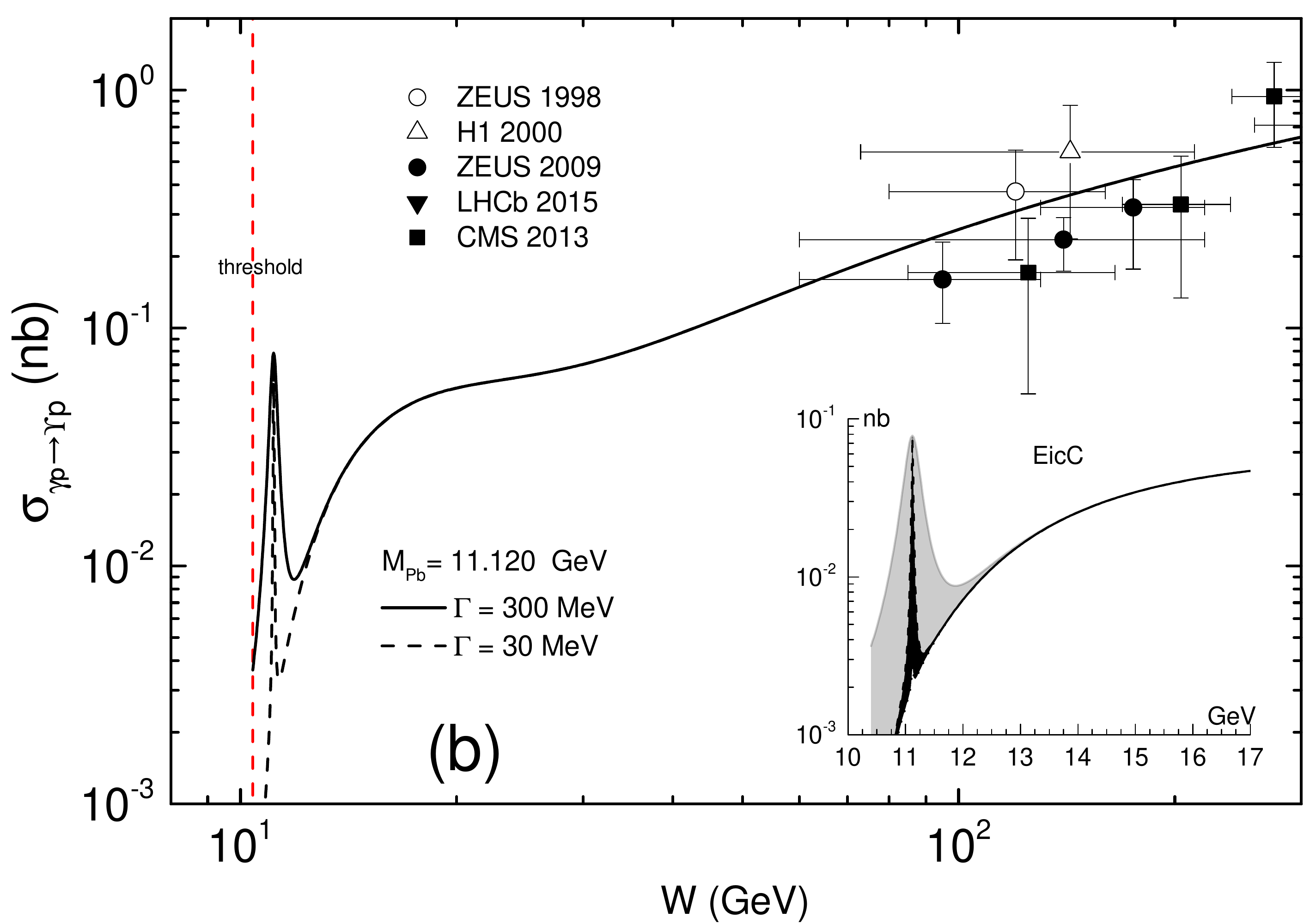}}
    \caption{The cross section of $\gamma p\to \Upsilon p$ with $\mathcal{B}(P_b \to \Upsilon p) = 5\%$ as a function of the  c.m. energy. The data are the same as those in Fig.~\ref{fig:Upsilon_all}. The inserted subplots enlarge the energy region covered by the proposed EicC. (a) The non-resonant contribution of the DVMP empirical formula is used. (b) The non-resonant contribution of the soft dipole Pomeron model is used. The bands in the inserted subplot on right bottom represent the range of $0.5\% < \mathcal{B}(P_b \to \Upsilon p) < 5\%$.
    \label{fig:addinUpsilon}}
  \end{center}
\end{figure}

The calculated results are presented in Fig.~\ref{fig:addinUpsilon}~(a) with $\mathcal{B}(P_b \to \Upsilon p) = 5\%$ and the non-resonant contribution of the DVMP empirical formula in Eq.~(\ref{eq:sigmaDVMP}). The background is smooth within the EicC energies in the range of 0.01 $\sim$ 0.02~nb. The peak cross section of the narrow $P_b$ is around 0.1~nb at most. The effects of both the narrow and wide $P_b$ are prominent, as can be seen. This is contrary to the decay of $\Lambda_b^0 \to K^- J/\psi p$ at LHCb, where a wide resonance is much harder to be identified due to the more complicated background. Notice that the DVMP parameterization does not consider the phase space, so that the results need to be considered as an upper limit in the low energy region as already mentioned.

We show the results with the non-resonant contribution of the soft dipole Pomeron model in Fig.~\ref{fig:addinUpsilon}~(b) with $0.5\% < \mathcal{B}(P_b \to \Upsilon p) < 5\%$. The background varies rapidly in the range of the EicC energies because of the phase space. The $P_b$ signal is still clearly visible if $\mathcal{B}(P_b \to \Upsilon p) > 1.0 \%$. It would be difficult to find the $P_b$ with $\mathcal{B}(P_b \to \Upsilon p)$ as small as 0.5\% in an unpolarized measurement, and therefore polarization observables are needed. The formalism for a detailed calculation of polarized measurements is well established~\cite{Wang:2015jsa,Wu:2019adv,Cao:2017njq}. But we will not pursue that aspect in this
paper, because the interference is definitely essential but out of control due to lack of low energy data. The $t$-dependence of the non-resonant contribution in the soft Pomeron model is very close to $e^{b\,t}$ with the same value for the slope $b$ in Eq.~(\ref{eq:2gluon}), because this slope is mainly driven by the data of the $J/\psi$ production in the soft dipole Pomeron model. It would be very interesting to look into the slope for the $\Upsilon$ once data are available in the future.

As shown in Fig.~\ref{fig:addinUpsilon}~(a), the non-resonant $\Upsilon$ photon-production at EicC energies is around 0.02~nb at most, and a reduction factor of about five is introduced by the two-body phase space, see Fig.~\ref{fig:addinUpsilon}~(b). The resonant $P_b$ photon-production in the peak energy is around 0.1~nb.  For reactions at electron ion colliders, a roughly two orders of magnitude smaller cross section is anticipated for the electroproduction comparing to above photon-production. Take the EicC as an example, about $5\times10^4$ signal events of  $ep\rightarrow e P_b \rightarrow  e \Upsilon p $ are expected with an integrated luminosity of 50 fb$^{-1}$. Even after considering the small leptonic decay branching fraction of the $\Upsilon$ and the detection efficiency, the observation of this channel is still optimistic at the EicC. The produced $P_b$ is not far away from the central rapidity region at the EicC energies, which is good for detection. A detailed simulation is under investigation and will be soon available for publication~\cite{CAO:2020EicC}.

{
The US-EIC project covers $ep$ c.m. energies of 30 $\sim$ 140 GeV (eRHIC), and its optimal energy is around 100 GeV~\cite{Accardi:2012qut,Boer:2011fh}.
It covers larger $Q^2$ range, and its electroproduction cross section of $\Upsilon p$ should be several times larger than that of EicC due to the much larger c.m. energies~\cite{CAO:2020EicC}.
Its designed luminosity is around one order of magnitude higher than that of EicC ($2 \sim 4 \times 10^{33}$ at EicC vs. $10^{34}$ or higher at US-EIC).
As a result, for the same duration of running time, the events produced at the US-EIC will be tens of times more than those  at EicC.
The final particles of the reaction of interest in EicC is within the middle rapidity range, while they are distributed in the large rapidity range at the US-EIC due to much higher c.m. energies, challenging the design of detector coverage for studying this spectroscopy issue. A detailed comparison of pentaquark electroproduction at EicC and other EICs with higher energies (US-EIC, LHeC) will be given in a forthcoming manuscript~\cite{Xie:2020pc}.
}

%
%%%%%%%%%%%%%%%%%%%%%%%%%%%%%%%%%%%%%%%%%%%%%%%%%%%%%%%%%%%%%%%%%%%%%%%%%%%%%%%%%%%%%%%%%%%%%%%%%%%%%%%%%%%%%%%%%%%%%%%
\section{Summary and Conclusion} \label{sec:summ}

In this paper, we made a detailed exploration of the non-resonant contribution to the $\gamma p \to \Upsilon p$, with the aim to find a reasonable estimation of the production rate at relatively low energies, where no data are available up to now. An extrapolation from the energies of the LHC and HERA data to low energies by several models gives us a reasonable estimate of the cross section below 100~GeV. We emphasize that this non-resonant contribution to the  $\gamma p \to \Upsilon p$ is related to several appealing topics.
It may give access to the $\Upsilon p$ scattering length, which is a key parameter for understanding of whether a bottomonium can be bound with the nucleon and light nuclei. Our results in Fig.~\ref{fig:Upsilon_all} in fact can be used to roughly estimate the scattering length, as done for that of $J/\psi p$~\cite{Strakovsky:2019bev}. It could be also decisive for extracting the information of the trace anomaly contribution to the nucleon mass, so finally solve the problem of the proton mass decomposition~\cite{Hatta:2018ina}.  We would like to further remark that the larger mass of the $\Upsilon$ than the $J/\psi$ could make it a better place for studying these issues, because the relative uncertainties of current quark mass and running coupling constants are smaller at high energies~\cite{Tanabashi:2018oca}
        \bea \nonumber
        \frac{ \Delta m_Q}{m_Q} \simeq
         \begin{cases}
         2.5\%  \\
         1.0\%
         \end{cases}, \qquad
        \frac{ \Delta \alpha_s}{\alpha_s} \simeq
         \begin{cases}
         7.8\%  &\mbox{for $J/\psi$}\\
         3.7\% &\mbox{for $\Upsilon$}
         \end{cases}.
        \eea
These issues may be notably clarified by measurements at the electron proton colliders.

After the study of the non-resonant contribution, we conducted a careful estimation of the production of $P_b$ in $\gamma p \to \Upsilon p$ under the assumption that the $P_b$ naturally inherits features from the $P_c$. If it is found in the photo- and electroproduction in the future, the $P_b$ will be firmly established as a genuine resonant state because resonant-like structures from triangle singularities are inapplicable to this reaction. We estimated the production yield at EIC machines based on the calculated cross sections and found that if the $\mathcal{B}(P_b \to \Upsilon p)$ is larger than 1.0\%, the $P_b$ states should be observed at the EicC through $\gamma p \to \Upsilon p$ process. On the other hand, if $\mathcal{B}(P_b \to \Upsilon p) $ is smaller than 1.0\% as predicted by Ref.~\cite{Lin:2018kcc}, then the $P_b$ states need to be searched for in the dominant decay channels $B^{(*)}\Lambda_b$ final states,
{whose branching fractions were predicted to be two to three orders of magnitude larger in Ref.~\cite{Lin:2018kcc}.
Then from Eq.~(\ref{eq:sigmaR}), the cross section of the $P_b$ photoproduction in $B^{(*)}\Lambda_b$ channel given by
\be
  \sigma_{P_b \to B^{(*)}\Lambda_b} = \frac{\mathcal{B}(P_b \to B^{(*)}\Lambda_b)}{\mathcal{B}(P_b \to \Upsilon p)} \sigma_{P_b \to \Upsilon p}
\ee
assuming the same production mechanism should also be larger by the same scale.
However, the cross section for the $\gamma p\to B^{(*)}\Lambda_b$ also receives a $t$-channel contribution from the exchange of a bottom meson. The calculation of such a contribution is highly model dependent because of the presence of the off-shell form factor.
%As a reference, the channel $\gamma p \to \bar D^{*0} \Lambda^+_c$ is calculated using an effective Lagrangian model in Ref.~\cite{Huang:2016tcr} which suggests the $t$-channel $D^0$ exchange to be dominant; however, the magnitude depends significantly on the cut-off in the form factor, ranging in several orders~\cite{Huang:2016tcr}.
As a reference, the channels $\gamma p \to \bar D^{0} \Lambda^+_c$ and $\gamma p \to \bar D^{*0} \Lambda^+_c$ are both calculated using an effective Lagrangian model in Ref.~\cite{Wu:2019adv} and the later is also computed with similar approach in Ref.~\cite{Huang:2016tcr}.
These models suggests the $t$-channel $D^0$ exchange to be dominant, however, the magnitude depends significantly on the cut-off in the form factor, ranging in several orders~\cite{Huang:2016tcr}.
The similar reaction in the bottom sector is anticipated to receive an even larger uncertainty due to the larger virtuality of the exchanged bottom meson.
Nevertheless, an estimation from the semi-inclusive $\gamma p \to b\bar{b}X$ would give a good guideline.}
Since the cross section of the semi-inclusive $\gamma p \to b\bar{b}X$ at high energies is found to be two orders of magnitude larger than that of the $\gamma p \to \Upsilon p$ by experiments~\cite{Luders:2001nb,Adloff:1999nr} and the next-to-leading order QCD calculation~\cite{Ellis:1988sb,Frixione:1994dv}, the open bottom channels $\bar B^{(*)}\Lambda_b$ from non-resonant contribution are expected to have larger cross section than that of $\Upsilon p$ channels.  So these $P_b$ states may be observed at EIC machines through the $\gamma p \to \bar B^{(*)}\Lambda_b$ reaction, if the detection efficiency of weak decay particles is promoted. In particular, almost all of the possible open bottom modes will have the $\bar B \Lambda_b$ in the final states. So a real measurement anticipated at EIC machines will surely clarify the situation to a large extent.

In brief, future EIC machines
can be used to search for the hidden-bottom pentaquark $P_b$ states, as the bottom partners of the $P_c$, in the $\gamma p \to \Upsilon p$ and open-bottom processes. New insights are expected into the physics of exotic hadrons. Also, the EicC can measure the cross section of $\gamma p \to \Upsilon p$ in relatively low energies, covering a variety of interesting physics aspects.

\begin{acknowledgments}

Useful discussions with Evgenij~Martynov and Nu Xu are gratefully acknowledged. We thank all the authors of the EicC white paper, especially Chao-Hsi Chang, Kuang-Ta Chao, Hongxia Huang, and Qiang Zhao for enlightening suggestions. This work was supported in part by the National Natural Science Foundation of China (Grant Nos. 11405222, 11621131001, 11835015, 11947302, 11975278, 11735003, 1191101002 and 11961141012), by the Thousand Talents Plan for Young Professionals, by the the Pioneer Hundred Talents Program of Chinese Academy of Sciences (CAS), by the Key Research Program of CAS (Grant No. XDPB09), by the CAS Key Research Program of Frontier Sciences (Grant No. QYZDB-SSW-SYS013), and by the CAS Center for Excellence in Particle Physics
(CCEPP).

\end{acknowledgments}


\begin{thebibliography}{1}

%\cite{Choi:2003ue}
\bibitem{Choi:2003ue}
  S.~K.~Choi {\it et al.} [Belle Collaboration],
  %``Observation of a narrow charmonium - like state in exclusive B+- ---> K+- pi+ pi- J / psi decays,''
  Phys.\ Rev.\ Lett.\  {\bf 91}, 262001 (2003)
  % doi:10.1103/PhysRevLett.91.262001
  [hep-ex/0309032].
  %%CITATION = doi:10.1103/PhysRevLett.91.262001;%%
  %1706 citations counted in INSPIRE as of 24 Dec 2019


%\cite{Chen:2016qju}
\bibitem{Chen:2016qju}
  H.-X.~Chen, W.~Chen, X.~Liu and S.-L.~Zhu,
  %``The hidden-charm pentaquark and tetraquark states,''
  Phys.\ Rept.\  {\bf 639}, 1 (2016)
  % doi:10.1016/j.physrep.2016.05.004
  [arXiv:1601.02092 [hep-ph]].
  %%CITATION = doi:10.1016/j.physrep.2016.05.004;%%
  %479 citations counted in INSPIRE as of 24 Dec 2019


%\cite{Guo:2017jvc}
\bibitem{Guo:2017jvc}
  F.-K.~Guo, C.~Hanhart, U.-G.~Mei{\ss}ner, Q.~Wang, Q.~Zhao and B.-S.~Zou,
  %``Hadronic molecules,''
  Rev.\ Mod.\ Phys.\  {\bf 90},  015004 (2018)
  % doi:10.1103/RevModPhys.90.015004
  [arXiv:1705.00141 [hep-ph]].
  %%CITATION = doi:10.1103/RevModPhys.90.015004;%%
  %327 citations counted in INSPIRE as of 24 Dec 2019


%\cite{Lebed:2016hpi}
\bibitem{Lebed:2016hpi}
  R.~F.~Lebed, R.~E.~Mitchell and E.~S.~Swanson,
  %``Heavy-Quark QCD Exotica,''
  Prog.\ Part.\ Nucl.\ Phys.\  {\bf 93}, 143 (2017)
  % doi:10.1016/j.ppnp.2016.11.003
  [arXiv:1610.04528 [hep-ph]].
  %%CITATION = doi:10.1016/j.ppnp.2016.11.003;%%
  %214 citations counted in INSPIRE as of 24 Dec 2019


%\cite{Esposito:2016noz}
\bibitem{Esposito:2016noz}
  A.~Esposito, A.~Pilloni and A.~D.~Polosa,
  %``Multiquark Resonances,''
  Phys.\ Rept.\  {\bf 668}, 1 (2017)
  % doi:10.1016/j.physrep.2016.11.002
  [arXiv:1611.07920 [hep-ph]].
  %%CITATION = doi:10.1016/j.physrep.2016.11.002;%%
  %244 citations counted in INSPIRE as of 24 Dec 2019


%\cite{Olsen:2017bmm}
\bibitem{Olsen:2017bmm}
  S.~L.~Olsen, T.~Skwarnicki and D.~Zieminska,
  %``Nonstandard heavy mesons and baryons: Experimental evidence,''
  Rev.\ Mod.\ Phys.\  {\bf 90},  015003 (2018)
  % doi:10.1103/RevModPhys.90.015003
  [arXiv:1708.04012 [hep-ph]].
  %%CITATION = doi:10.1103/RevModPhys.90.015003;%%
  %202 citations counted in INSPIRE as of 24 Dec 2019


%\cite{Liu:2019zoy}
\bibitem{Liu:2019zoy}
  Y.-R.~Liu, H.-X.~Chen, W.~Chen, X.~Liu and S.-L.~Zhu,
  %``Pentaquark and Tetraquark states,''
  Prog.\ Part.\ Nucl.\ Phys.\  {\bf 107}, 237 (2019)
  % doi:10.1016/j.ppnp.2019.04.003
  [arXiv:1903.11976 [hep-ph]].
  %%CITATION = doi:10.1016/j.ppnp.2019.04.003;%%
  %63 citations counted in INSPIRE as of 24 Dec 2019

\bibitem{Brambilla:2019esw}
  N.~Brambilla, S.~Eidelman, C.~Hanhart, A.~Nefediev, C.~P.~Shen, C.~E.~Thomas, A.~Vairo and C.~Z.~Yuan,
  %``The $XYZ$ states: experimental and theoretical status and perspectives,''
  arXiv:1907.07583 [hep-ex].

%\cite{Guo:2019twa}
\bibitem{Guo:2019twa}
  F.~K.~Guo, X.~H.~Liu and S.~Sakai,
  %``Threshold cusps and triangle singularities in hadronic reactions,''
  Prog.\ Part.\ Nucl.\ Phys.\  {\bf 112}, 103757 (2020)
%  doi:10.1016/j.ppnp.2020.103757
  [arXiv:1912.07030 [hep-ph]].
  %%CITATION = doi:10.1016/j.ppnp.2020.103757;%%
  %20 citations counted in INSPIRE as of 14 Apr 2020

%\cite{Ablikim:2013mio}
\bibitem{Ablikim:2013mio}
  M.~Ablikim {\it et al.} [BESIII Collaboration],
  %``Observation of a Charged Charmoniumlike Structure in $e^+e^-$ �� $��^+��^-$ J/�� at $\sqrt{s}$ =4.26  GeV,''
  Phys.\ Rev.\ Lett.\  {\bf 110}, 252001 (2013)
  % doi:10.1103/PhysRevLett.110.252001
  [arXiv:1303.5949 [hep-ex]].
  %%CITATION = doi:10.1103/PhysRevLett.110.252001;%%
  %705 citations counted in INSPIRE as of 24 Dec 2019


%\cite{Liu:2013dau}
\bibitem{Liu:2013dau}
  Z.~Q.~Liu {\it et al.} [Belle Collaboration],
  %``Study of $e^+e^- �� ��^+ ��^- J/��$ and Observation of a Charged Charmoniumlike State at Belle,''
  Phys.\ Rev.\ Lett.\  {\bf 110}, 252002 (2013)
  Erratum: [Phys.\ Rev.\ Lett.\  {\bf 111}, 019901 (2013)]
  % doi:10.1103/PhysRevLett.110.252002, 10.1103/PhysRevLett.111.019901
  [arXiv:1304.0121 [hep-ex]].
  %%CITATION = doi:10.1103/PhysRevLett.110.252002, 10.1103/PhysRevLett.111.019901;%%
  %600 citations counted in INSPIRE as of 24 Dec 2019


%\cite{Ablikim:2013wzq}
\bibitem{Ablikim:2013wzq}
  M.~Ablikim {\it et al.} [BESIII Collaboration],
  %``Observation of a Charged Charmoniumlike Structure $Z_c$(4020) and Search for the $Z_c$(3900) in $e^+e^- \to ��^+��^-h_c$,''
  Phys.\ Rev.\ Lett.\  {\bf 111},  242001 (2013)
  % doi:10.1103/PhysRevLett.111.242001
  [arXiv:1309.1896 [hep-ex]].
  %%CITATION = doi:10.1103/PhysRevLett.111.242001;%%
  %347 citations counted in INSPIRE as of 24 Dec 2019


%\cite{Belle:2011aa}
\bibitem{Belle:2011aa}
  A.~Bondar {\it et al.} [Belle Collaboration],
  %``Observation of two charged bottomonium-like resonances in Y(5S) decays,''
  Phys.\ Rev.\ Lett.\  {\bf 108}, 122001 (2012)
  % doi:10.1103/PhysRevLett.108.122001
  [arXiv:1110.2251 [hep-ex]].
  %%CITATION = doi:10.1103/PhysRevLett.108.122001;%%
  %517 citations counted in INSPIRE as of 24 Dec 2019


%\cite{Collaboration:2017njt}
\bibitem{Collaboration:2017njt}
  M.~Ablikim {\it et al.} [BESIII Collaboration],
  %``Determination of the Spin and Parity of the $Z_c(3900)$,''
  Phys.\ Rev.\ Lett.\  {\bf 119},  072001 (2017)
  % doi:10.1103/PhysRevLett.119.072001
  [arXiv:1706.04100 [hep-ex]].
  %%CITATION = doi:10.1103/PhysRevLett.119.072001;%%
  %49 citations counted in INSPIRE as of 24 Dec 2019


%\cite{Garmash:2014dhx}
\bibitem{Garmash:2014dhx}
  A.~Garmash {\it et al.} [Belle Collaboration],
  %``Amplitude analysis of $e^+e^- \to \Upsilon(nS) \pi^+\pi^-$ at $\sqrt{s}=10.865$~GeV,''
  Phys.\ Rev.\ D {\bf 91},  072003 (2015)
  % doi:10.1103/PhysRevD.91.072003
  [arXiv:1403.0992 [hep-ex]].
  %%CITATION = doi:10.1103/PhysRevD.91.072003;%%
  %69 citations counted in INSPIRE as of 24 Dec 2019


%\cite{Ablikim:2018ofc}
\bibitem{Ablikim:2018ofc}
  M.~Ablikim {\it et al.} [BESIII Collaboration],
  %``Search for a strangeonium-like structure $Z_s$ decaying into $\phi \pi$ and a measurement of the cross section $e^+e^-\rightarrow\phi\pi\pi$,''
  Phys.\ Rev.\ D {\bf 99},  011101 (2019)
  % doi:10.1103/PhysRevD.99.011101
  [arXiv:1801.10384 [hep-ex]].
  %%CITATION = doi:10.1103/PhysRevD.99.011101;%%
  %2 citations counted in INSPIRE as of 24 Dec 2019


%\cite{Huang:2018ehi}
\bibitem{Huang:2018ehi}
  H.~Huang, X.~Zhu and J.~Ping,
  %``$P_{c}$-like pentaquarks in hidden strange sector,''
  Phys.\ Rev.\ D {\bf 97},  094019 (2018)
  % doi:10.1103/PhysRevD.97.094019
  [arXiv:1803.05267 [hep-ph]].
  %%CITATION = doi:10.1103/PhysRevD.97.094019;%%
  %8 citations counted in INSPIRE as of 24 Dec 2019


%\cite{Liu:2018nse}
\bibitem{Liu:2018nse}
  X.~Liu, H.~Huang and J.~Ping,
  %``Hidden strange pentaquark states in constituent quark models,''
  Phys.\ Rev.\ C {\bf 98},  055203 (2018)
  % doi:10.1103/PhysRevC.98.055203
  [arXiv:1807.03195 [hep-ph]].
  %%CITATION = doi:10.1103/PhysRevC.98.055203;%%

  \bibitem{Gao:2000az}
  H.~Gao, T.~S.~H.~Lee and V.~Marinov,
  %``Phi0 - N bound state,''
  Phys.\ Rev.\ C {\bf 63}, 022201 (2001)
  % doi:10.1103/PhysRevC.63.022201
  [nucl-th/0010042].

%\cite{He:2018plt}
\bibitem{He:2018plt}
  J.~He, H.~Huang, D.-Y.~Chen and X.~Zhu,
  %``Hidden-strange molecular states and the N? bound states via a QCD van der Waals force,''
  Phys.\ Rev.\ D {\bf 98},  094019 (2018)
  % doi:10.1103/PhysRevD.98.094019
  [arXiv:1804.09383 [hep-ph]].
  %%CITATION = doi:10.1103/PhysRevD.98.094019;%%
  %3 citations counted in INSPIRE as of 24 Dec 2019


%\cite{Cao:2017njq}
\bibitem{Cao:2017njq}
  X.~Cao and H.~Lenske,
  %``Compton scattering off proton in the third resonance region,''
  Phys.\ Lett.\ B {\bf 772}, 274 (2017)
  % doi:10.1016/j.physletb.2017.06.063
  [arXiv:1702.02692 [nucl-th]].
  %%CITATION = doi:10.1016/j.physletb.2017.06.063;%%
  %2 citations counted in INSPIRE as of 24 Dec 2019


%\cite{Cao:2014vma}
\bibitem{Cao:2014vma}
  X.~Cao,
  %``Disentangling the nature of resonances in coupled-channel models,''
  Chin.\ Phys.\ C {\bf 39},  041002 (2015)
  % doi:10.1088/1674-1137/39/4/041002
  [arXiv:1404.6651 [hep-ph]].
  %%CITATION = doi:10.1088/1674-1137/39/4/041002;%%
  %1 citations counted in INSPIRE as of 24 Dec 2019


%\cite{Cao:2018vmv}
\bibitem{Cao:2018vmv}
  X.~Cao and J.~P.~Dai,
  %``Spin parity of  $Z_c^-$(4100), $Z_1^+$(4050) and $Z_2^+$(4250),''
  Phys.\ Rev.\ D {\bf 100},  054004 (2019)
  % doi:10.1103/PhysRevD.100.054004
  [arXiv:1811.06434 [hep-ph]].
  %%CITATION = doi:10.1103/PhysRevD.100.054004;%%
  %5 citations counted in INSPIRE as of 24 Dec 2019

%\cite{Cao:2014mea}
\bibitem{Cao:2014mea}
  X.~Cao and J.-J.~Xie,
  %``Nucleon resonances in $\pi$N $\to$ �ǡ�N and J/$\psi \to p\overline p$�ǡ�$^*$,''
  Chin.\ Phys.\ C {\bf 40},  083103 (2016)
  % doi:10.1088/1674-1137/40/8/083103
  [arXiv:1411.1493 [nucl-th]].
  %%CITATION = doi:10.1088/1674-1137/40/8/083103;%%
  %1 citations counted in INSPIRE as of 24 Dec 2019

\bibitem{Lebed:2015dca}
  R.~F.~Lebed,
  %``Do the $P_c^+$ pentaquarks have strange siblings?,''
  Phys.\ Rev.\ D {\bf 92},  114030 (2015)
  % doi:10.1103/PhysRevD.92.114030
  [arXiv:1510.06648 [hep-ph]].


%\cite{He:2017aps}
\bibitem{He:2017aps}
  J.~He,
  %``Nucleon resonances $N(1875)$ and $N(2100)$ as strange partners of LHCb pentaquarks,''
  Phys.\ Rev.\ D {\bf 95},  074031 (2017)
  % doi:10.1103/PhysRevD.95.074031
  [arXiv:1701.03738 [hep-ph]].
  %%CITATION = doi:10.1103/PhysRevD.95.074031;%%
  %21 citations counted in INSPIRE as of 24 Dec 2019


%\cite{An:2018vmk}
\bibitem{An:2018vmk}
  C.-S.~An, J.-J.~Xie and G.~Li,
  %``Decay patterns of low-lying $N_{s\bar{s}}$ states to the strangeness channels,''
  Phys.\ Rev.\ C {\bf 98},  045201 (2018)
  % doi:10.1103/PhysRevC.98.045201
  [arXiv:1809.04934 [hep-ph]].
  %%CITATION = doi:10.1103/PhysRevC.98.045201;%%
  %1 citations counted in INSPIRE as of 24 Dec 2019


%\cite{Gao:2017hya}
\bibitem{Gao:2017hya}
  H.~Gao, H.~Huang, T.~Liu, J.~Ping, F.~Wang and Z.~Zhao,
  %``Search for a hidden strange baryon-meson bound state from ? production in a nuclear medium,''
  Phys.\ Rev.\ C {\bf 95},  055202 (2017)
  % doi:10.1103/PhysRevC.95.055202
  [arXiv:1701.03210 [hep-ph]].
  %%CITATION = doi:10.1103/PhysRevC.95.055202;%%
  %17 citations counted in INSPIRE as of 24 Dec 2019


%\cite{Mibe:2005er}
\bibitem{Mibe:2005er}
  T.~Mibe {\it et al.} [LEPS Collaboration],
  %``Diffractive phi-meson photoproduction on proton near threshold,''
  Phys.\ Rev.\ Lett.\  {\bf 95}, 182001 (2005)
  % doi:10.1103/PhysRevLett.95.182001
  [nucl-ex/0506015].
  %%CITATION = doi:10.1103/PhysRevLett.95.182001;%%
  %104 citations counted in INSPIRE as of 24 Dec 2019


%\cite{Kiswandhi:2010ub}
\bibitem{Kiswandhi:2010ub}
  A.~Kiswandhi, J.-J.~Xie and S.-N.~Yang,
  %``Is the nonmonotonic behavior in the cross section of phi photoproduction near threshold a signature of a resonance?,''
  Phys.\ Lett.\ B {\bf 691}, 214 (2010)
  % doi:10.1016/j.physletb.2010.06.029
  [arXiv:1005.2105 [hep-ph]].
  %%CITATION = doi:10.1016/j.physletb.2010.06.029;%%
  %23 citations counted in INSPIRE as of 24 Dec 2019


%\cite{Kiswandhi:2011cq}
\bibitem{Kiswandhi:2011cq}
  A.~Kiswandhi and S.-N.~Yang,
  %``On the near-threshold peak structure in the differential cross section of \phi-meson photoproduction: indication of a missing resonance with non-negligible strangeness content,''
  Phys.\ Rev.\ C {\bf 86}, 015203 (2012)
  Erratum: [Phys.\ Rev.\ C {\bf 86}, 019904 (2012)]
  % doi:10.1103/PhysRevC.86.019904, 10.1103/PhysRevC.86.015203
  [arXiv:1112.6105 [nucl-th]].
  %%CITATION = doi:10.1103/PhysRevC.86.019904, 10.1103/PhysRevC.86.015203;%%
  %22 citations counted in INSPIRE as of 24 Dec 2019

\bibitem{Pal:2017ypp}
  B.~Pal {\it et al.} [Belle Collaboration],
  %``Search for $\Lambda_c^+\to\phi p \pi^0$ and branching fraction measurement of $\Lambda_c^+\to K^-\pi^+ p \pi^0$,''
  Phys.\ Rev.\ D {\bf 96},  051102 (2017)
  % doi:10.1103/PhysRevD.96.051102
  [arXiv:1707.00089 [hep-ex]].


%\cite{Xie:2017mbe}
\bibitem{Xie:2017mbe}
  J.-J.~Xie and F.-K.~Guo,
  %``Triangular singularity and a possible $\phi p$ resonance in the $\Lambda^+_c \to \pi^0 \phi p$ decay,''
  Phys.\ Lett.\ B {\bf 774}, 108 (2017)
  % doi:10.1016/j.physletb.2017.09.060
  [arXiv:1709.01416 [hep-ph]].
  %%CITATION = doi:10.1016/j.physletb.2017.09.060;%%
  %19 citations counted in INSPIRE as of 24 Dec 2019

%\cite{Aaij:2015tga}
\bibitem{Aaij:2015tga}
  R.~Aaij {\it et al.} [LHCb Collaboration],
  %``Observation of $J/\psi p$ Resonances Consistent with Pentaquark States in $\Lambda_b^0 \to J/\psi K^- p$ Decays,''
  Phys.\ Rev.\ Lett.\  {\bf 115}, 072001 (2015)
  % doi:10.1103/PhysRevLett.115.072001
  [arXiv:1507.03414 [hep-ex]].
  %%CITATION = doi:10.1103/PhysRevLett.115.072001;%%
  %900 citations counted in INSPIRE as of 24 Dec 2019


%\cite{Aaij:2019vzc}
\bibitem{Aaij:2019vzc}
  R.~Aaij {\it et al.} [LHCb Collaboration],
  %``Observation of a narrow pentaquark state, $P_c(4312)^+$, and of two-peak structure of the $P_c(4450)^+$,''
  Phys.\ Rev.\ Lett.\  {\bf 122},  222001 (2019)
  % doi:10.1103/PhysRevLett.122.222001
  [arXiv:1904.03947 [hep-ex]].
  %%CITATION = doi:10.1103/PhysRevLett.122.222001;%%
  %115 citations counted in INSPIRE as of 24 Dec 2019


%\cite{Wu:2010jy}
\bibitem{Wu:2010jy}
  J.-J.~Wu, R.~Molina, E.~Oset and B.-S.~Zou,
  %``Prediction of narrow $N^*$ and $\Lambda^*$ resonances with hidden charm above 4 GeV,''
  Phys.\ Rev.\ Lett.\  {\bf 105}, 232001 (2010)
  % doi:10.1103/PhysRevLett.105.232001
  [arXiv:1007.0573 [nucl-th]].
  %%CITATION = doi:10.1103/PhysRevLett.105.232001;%%
  %242 citations counted in INSPIRE as of 24 Dec 2019


%\cite{Wu:2010vk}
\bibitem{Wu:2010vk}
  J.-J.~Wu, R.~Molina, E.~Oset and B.-S.~Zou,
  %``Dynamically generated $N^{*}$ and $\Lambda^*$ resonances in the hidden charm sector around 4.3 GeV,''
  Phys.\ Rev.\ C {\bf 84}, 015202 (2011)
  % doi:10.1103/PhysRevC.84.015202
  [arXiv:1011.2399 [nucl-th]].
  %%CITATION = doi:10.1103/PhysRevC.84.015202;%%
  %159 citations counted in INSPIRE as of 24 Dec 2019


%\cite{Wang:2011rga}
\bibitem{Wang:2011rga}
  W.-L.~Wang, F.~Huang, Z.-Y.~Zhang and B.-S.~Zou,
  %``$\Sigma_c \bar{D}$ and $\Lambda_c \bar{D}$ states in a chiral quark model,''
  Phys.\ Rev.\ C {\bf 84}, 015203 (2011)
  % doi:10.1103/PhysRevC.84.015203
  [arXiv:1101.0453 [nucl-th]].
  %%CITATION = doi:10.1103/PhysRevC.84.015203;%%
  %92 citations counted in INSPIRE as of 24 Dec 2019


%\cite{Yang:2011wz}
\bibitem{Yang:2011wz}
  Z.-C.~Yang, Z.-F.~Sun, J.~He, X.~Liu and S.-L.~Zhu,
  %``The possible hidden-charm molecular baryons composed of anti-charmed meson and charmed baryon,''
  Chin.\ Phys.\ C {\bf 36}, 6 (2012)
  % doi:10.1088/1674-1137/36/1/002, 10.1088/1674-1137/36/3/006
  [arXiv:1105.2901 [hep-ph]].
  %%CITATION = doi:10.1088/1674-1137/36/1/002, 10.1088/1674-1137/36/3/006;%%
  %131 citations counted in INSPIRE as of 24 Dec 2019


%\cite{Wang:2015jsa}
\bibitem{Wang:2015jsa}
  Q.~Wang, X.-H.~Liu and Q.~Zhao,
  %``Photoproduction of hidden charm pentaquark states $P_c^+(4380)$ and $P_c^+(4450)$,''
  Phys.\ Rev.\ D {\bf 92}, 034022 (2015)
  % doi:10.1103/PhysRevD.92.034022
  [arXiv:1508.00339 [hep-ph]].
  %%CITATION = doi:10.1103/PhysRevD.92.034022;%%
  %103 citations counted in INSPIRE as of 24 Dec 2019


%\cite{Kubarovsky:2015aaa}
\bibitem{Kubarovsky:2015aaa}
  V.~Kubarovsky and M.~B.~Voloshin,
  %``Formation of hidden-charm pentaquarks in photon-nucleon collisions,''
  Phys.\ Rev.\ D {\bf 92},  031502 (2015)
  % doi:10.1103/PhysRevD.92.031502
  [arXiv:1508.00888 [hep-ph]].
  %%CITATION = doi:10.1103/PhysRevD.92.031502;%%
  %120 citations counted in INSPIRE as of 24 Dec 2019



%\cite{Karliner:2015voa}
\bibitem{Karliner:2015voa}
  M.~Karliner and J.~L.~Rosner,
  %``Photoproduction of Exotic Baryon Resonances,''
  Phys.\ Lett.\ B {\bf 752}, 329 (2016)
  % doi:10.1016/j.physletb.2015.11.068
  [arXiv:1508.01496 [hep-ph]].
  %%CITATION = doi:10.1016/j.physletb.2015.11.068;%%
  %87 citations counted in INSPIRE as of 24 Dec 2019



%\cite{Guo:2015umn}
\bibitem{Guo:2015umn}
  F.-K.~Guo, U.-G.~Mei{\ss}ner, W.~Wang and Z.~Yang,
  %``How to reveal the exotic nature of the P$_c$(4450),''
  Phys.\ Rev.\ D {\bf 92},  071502 (2015)
  % doi:10.1103/PhysRevD.92.071502
  [arXiv:1507.04950 [hep-ph]].
  %%CITATION = doi:10.1103/PhysRevD.92.071502;%%
  %224 citations counted in INSPIRE as of 24 Dec 2019

%\cite{Liu:2015fea}
\bibitem{Liu:2015fea}
  X.-H.~Liu, Q.~Wang and Q.~Zhao,
  %``Understanding the newly observed heavy pentaquark candidates,''
  Phys.\ Lett.\ B {\bf 757}, 231 (2016)
  % doi:10.1016/j.physletb.2016.03.089
  [arXiv:1507.05359 [hep-ph]].
  %%CITATION = doi:10.1016/j.physletb.2016.03.089;%%
  %176 citations counted in INSPIRE as of 25 Dec 2019

%\cite{Guo:2016bkl}
\bibitem{Guo:2016bkl}
  F.-K.~Guo, U.-G.~Mei{\ss}ner, J.~Nieves and Z.~Yang,
  %``Remarks on the $P_c$ structures and triangle singularities,''
  Eur.\ Phys.\ J.\ A {\bf 52},  318 (2016)
  % doi:10.1140/epja/i2016-16318-4
  [arXiv:1605.05113 [hep-ph]].
  %%CITATION = doi:10.1140/epja/i2016-16318-4;%%
  %53 citations counted in INSPIRE as of 24 Dec 2019

\bibitem{Bayar:2016ftu}
  M.~Bayar, F.~Aceti, F.-K.~Guo and E.~Oset,
  %``A Discussion on Triangle Singularities in the $\Lambda_b \to J/\psi K^{-} p$ Reaction,''
  Phys.\ Rev.\ D {\bf 94}, 074039 (2016)
  % doi:10.1103/PhysRevD.94.074039
  [arXiv:1609.04133 [hep-ph]].

%\cite{Liu:2019dqc}
\bibitem{Liu:2019dqc}
  X.~H.~Liu, G.~Li, J.-J.~Xie and Q.~Zhao,
  %``Visible narrow cusp structure in $\Lambda_c^+\to p K^- \pi^+$ enhanced by triangle singularity,''
  Phys.\ Rev.\ D {\bf 100},  054006 (2019)
  % doi:10.1103/PhysRevD.100.054006
  [arXiv:1906.07942 [hep-ph]].
  %%CITATION = doi:10.1103/PhysRevD.100.054006;%%
  %4 citations counted in INSPIRE as of 25 Dec 2019

%\cite{Ali:2019lzf}
\bibitem{Ali:2019lzf}
  A.~Ali {\it et al.} [GlueX Collaboration],
  %``First Measurement of Near-Threshold J/�� Exclusive Photoproduction off the Proton,''
  Phys.\ Rev.\ Lett.\  {\bf 123},  072001 (2019)
  % doi:10.1103/PhysRevLett.123.072001
  [arXiv:1905.10811 [nucl-ex]].
  %%CITATION = doi:10.1103/PhysRevLett.123.072001;%%
  %26 citations counted in INSPIRE as of 24 Dec 2019


%\cite{Cao:2019kst}
\bibitem{Cao:2019kst}
  X.~Cao and J.~p.~Dai,
  %``Confronting pentaquark photoproduction with new LHCb observations,''
  Phys.\ Rev.\ D {\bf 100},  054033 (2019)
  % doi:10.1103/PhysRevD.100.054033
  [arXiv:1904.06015 [hep-ph]].
  %%CITATION = doi:10.1103/PhysRevD.100.054033;%%
  %22 citations counted in INSPIRE as of 24 Dec 2019


%\cite{Winney:2019edt}
\bibitem{Winney:2019edt}
  D.~Winney {\it et al.} [JPAC Collaboration],
  %``Double polarization observables in pentaquark photoproduction,''
  Phys.\ Rev.\ D {\bf 100},  034019 (2019)
  % doi:10.1103/PhysRevD.100.034019
  [arXiv:1907.09393 [hep-ph]].
  %%CITATION = doi:10.1103/PhysRevD.100.034019;%%
  %6 citations counted in INSPIRE as of 24 Dec 2019

\bibitem{Chen:2019asm}
  R.~Chen, Z.~F.~Sun, X.~Liu and S.~L.~Zhu,
  %``Strong LHCb evidence supporting the existence of the hidden-charm molecular pentaquarks,''
  Phys.\ Rev.\ D {\bf 100}, 011502 (2019)
  % doi:10.1103/PhysRevD.100.011502
  [arXiv:1903.11013 [hep-ph]].

%\cite{Guo:2019fdo}
\bibitem{Guo:2019fdo}
  F.-K.~Guo, H.-J.~Jing, U.-G.~Mei{\ss}ner and S.~Sakai,
  %``Isospin breaking decays as a diagnosis of the hadronic molecular structure of the $P_c(4457)$,''
  Phys.\ Rev.\ D {\bf 99},  091501 (2019)
  % doi:10.1103/PhysRevD.99.091501
  [arXiv:1903.11503 [hep-ph]].
  %%CITATION = doi:10.1103/PhysRevD.99.091501;%%
  %45 citations counted in INSPIRE as of 24 Dec 2019


%\cite{Guo:2019kdc}
\bibitem{Guo:2019kdc}
  Z.-H.~Guo and J.~A.~Oller,
  %``Anatomy of the newly observed hidden-charm pentaquark states: $P_c(4312)$, $P_c(4440)$ and $P_c(4457)$,''
  Phys.\ Lett.\ B {\bf 793}, 144 (2019)
  % doi:10.1016/j.physletb.2019.04.053
  [arXiv:1904.00851 [hep-ph]].
  %%CITATION = doi:10.1016/j.physletb.2019.04.053;%%
  %40 citations counted in INSPIRE as of 24 Dec 2019

\bibitem{Eides:2019tgv}
  M.~I.~Eides, V.~Y.~Petrov and M.~V.~Polyakov,
  %``New LHCb pentaquarks as hadrocharmonium states,''
  arXiv:1904.11616 [hep-ph].

%\cite{Wang:2019got}
\bibitem{Wang:2019got}
  Z.~G.~Wang,
  %``Analysis of the $P_c(4312)$, $P_c(4440)$, $P_c(4457)$ and related hidden-charm pentaquark states with QCD sum rules,''
  Int.\ J.\ Mod.\ Phys.\ A {\bf 35}, no. 01, 2050003 (2020)
%  doi:10.1142/S0217751X20500037
  [arXiv:1905.02892 [hep-ph]].
  %%CITATION = doi:10.1142/S0217751X20500037;%%
  %21 citations counted in INSPIRE as of 10 Mar 2020

\bibitem{Ali:2019clg}
  A.~Ali, I.~Ahmed, M.~J.~Aslam, A.~Y.~Parkhomenko and A.~Rehman,
  %``Mass spectrum of the hidden-charm pentaquarks in the compact diquark model,''
  JHEP {\bf 1910}, 256 (2019)
  % doi:10.1007/JHEP10(2019)256
  [arXiv:1907.06507 [hep-ph]].

\bibitem{Burns:2019iih}
  T.~J.~Burns and E.~S.~Swanson,
  %``Molecular Interpretation of the $P_c(4440)$ and $P_c(4457)$ States,''
  Phys.\ Rev.\ D {\bf 100}, 114033 (2019)
  % doi:10.1103/PhysRevD.100.114033
  [arXiv:1908.03528 [hep-ph]].

%\cite{Liu:2019tjn}
\bibitem{Liu:2019tjn}
  M.-Z.~Liu, Y.-W.~Pan, F.-Z.~Peng, M.~S{\'a}nchez S{\'a}nchez, L.-S.~Geng, A.~Hosaka and M.~Pavon Valderrama,
  %``Emergence of a complete heavy-quark spin symmetry multiplet: seven molecular pentaquarks in light of the latest LHCb analysis,''
  Phys.\ Rev.\ Lett.\  {\bf 122},  242001 (2019)
  % doi:10.1103/PhysRevLett.122.242001
  [arXiv:1903.11560 [hep-ph]].
  %%CITATION = doi:10.1103/PhysRevLett.122.242001;%%
  %59 citations counted in INSPIRE as of 24 Dec 2019


%\cite{Xiao:2019aya}
\bibitem{Xiao:2019aya}
  C.-W.~Xiao, J.~Nieves and E.~Oset,
  %``Heavy quark spin symmetric molecular states from ${\bar D}^{(*)}\Sigma_c^{(*)}$ and other coupled channels in the light of the recent LHCb pentaquarks,''
  Phys.\ Rev.\ D {\bf 100},  014021 (2019)
  % doi:10.1103/PhysRevD.100.014021
  [arXiv:1904.01296 [hep-ph]].
  %%CITATION = doi:10.1103/PhysRevD.100.014021;%%
  %42 citations counted in INSPIRE as of 24 Dec 2019

%\cite{Du:2019pij}
\bibitem{Du:2019pij}
  M.~L.~Du, V.~Baru, F.~K.~Guo, C.~Hanhart, U.~G.~Mei{\ss}ner, J.~A.~Oller and Q.~Wang,
  %``Interpretation of the LHCb $P_c$ States as Hadronic Molecules and Hints of a Narrow $P_c(4380)$,''
  Phys.\ Rev.\ Lett.\  {\bf 124}, no. 7, 072001 (2020)
%  doi:10.1103/PhysRevLett.124.072001
  [arXiv:1910.11846 [hep-ph]].
  %%CITATION = doi:10.1103/PhysRevLett.124.072001;%%
  %8 citations counted in INSPIRE as of 10 Mar 2020

%\cite{Wu:2010rv}
\bibitem{Wu:2010rv}
  J.-J.~Wu, L. Zhao and B.-S.~Zou,
  %``Prediction of super-heavy $N^*$ and $\Lambda^*$ resonances with hidden beauty,''
  Phys.\ Lett.\ B {\bf 709}, 70 (2012)
  % doi:10.1016/j.physletb.2012.01.068
  [arXiv:1011.5743 [hep-ph]].
  %%CITATION = doi:10.1016/j.physletb.2012.01.068;%%
  %64 citations counted in INSPIRE as of 24 Dec 2019


%\cite{Xiao:2013jla}
\bibitem{Xiao:2013jla}
  C.-W.~Xiao and E.~Oset,
  %``Hidden beauty baryon states in the local hidden gauge approach with heavy quark spin symmetry,''
  Eur.\ Phys.\ J.\ A {\bf 49}, 139 (2013)
  % doi:10.1140/epja/i2013-13139-y
  [arXiv:1305.0786 [hep-ph]].
  %%CITATION = doi:10.1140/epja/i2013-13139-y;%%
  %25 citations counted in INSPIRE as of 24 Dec 2019

%\cite{Karliner:2015ina}
\bibitem{Karliner:2015ina}
  M.~Karliner and J.~L.~Rosner,
  %``New Exotic Meson and Baryon Resonances from Doubly-Heavy Hadronic Molecules,''
  Phys.\ Rev.\ Lett.\  {\bf 115},  122001 (2015)
  % doi:10.1103/PhysRevLett.115.122001
  [arXiv:1506.06386 [hep-ph]].
  %%CITATION = doi:10.1103/PhysRevLett.115.122001;%%
  %150 citations counted in INSPIRE as of 24 Dec 2019


%\cite{Breitweg:1998ki}
\bibitem{Breitweg:1998ki}
  J.~Breitweg {\it et al.} [ZEUS Collaboration],
  %``Measurement of elastic Upsilon photoproduction at HERA,''
  Phys.\ Lett.\ B {\bf 437}, 432 (1998)
  % doi:10.1016/S0370-2693(98)01081-8
  [hep-ex/9807020].
  %%CITATION = doi:10.1016/S0370-2693(98)01081-8;%%
  %141 citations counted in INSPIRE as of 24 Dec 2019


%\cite{Chekanov:2009zz}
\bibitem{Chekanov:2009zz}
  S.~Chekanov {\it et al.} [ZEUS Collaboration],
  %``Exclusive photoproduction of upsilon mesons at HERA,''
  Phys.\ Lett.\ B {\bf 680}, 4 (2009)
  % doi:10.1016/j.physletb.2009.07.066
  [arXiv:0903.4205 [hep-ex]].
  %%CITATION = doi:10.1016/j.physletb.2009.07.066;%%
  %96 citations counted in INSPIRE as of 24 Dec 2019


%\cite{Adloff:2000vm}
\bibitem{Adloff:2000vm}
  C.~Adloff {\it et al.} [H1 Collaboration],
  %``Elastic photoproduction of J / psi and Upsilon mesons at HERA,''
  Phys.\ Lett.\ B {\bf 483}, 23 (2000)
  % doi:10.1016/S0370-2693(00)00530-X
  [hep-ex/0003020].
  %%CITATION = doi:10.1016/S0370-2693(00)00530-X;%%
  %255 citations counted in INSPIRE as of 24 Dec 2019


%\cite{CMS:2016nct}
\bibitem{CMS:2016nct}
  CMS Collaboration [CMS Collaboration],
  %``Measurement of exclusive Y photoproduction in pPb collisions at $\sqrt{s_{_\mathrm{NN}}} = 5.02~\mathrm{TeV}$,''
  CMS-PAS-FSQ-13-009.
  %%CITATION = CMS-PAS-FSQ-13-009;%%
  %15 citations counted in INSPIRE as of 24 Dec 2019

%\cite{Aaij:2015kea}
\bibitem{Aaij:2015kea}
  R.~Aaij {\it et al.} [LHCb Collaboration],
  %``Measurement of the exclusive �� production cross-section in pp collisions at $ \sqrt{s}=7 $ TeV and 8 TeV,''
  JHEP {\bf 1509}, 084 (2015)
  % doi:10.1007/JHEP09(2015)084
  [arXiv:1505.08139 [hep-ex]].
  %%CITATION = doi:10.1007/JHEP09(2015)084;%%
  %107 citations counted in INSPIRE as of 24 Dec 2019



%\cite{Favart:2015umi}
\bibitem{Favart:2015umi}
  L.~Favart, M.~Guidal, T.~Horn and P.~Kroll,
  %``Deeply Virtual Meson Production on the nucleon,''
  Eur.\ Phys.\ J.\ A {\bf 52},  158 (2016)
  % doi:10.1140/epja/i2016-16158-2
  [arXiv:1511.04535 [hep-ph]].
  %%CITATION = doi:10.1140/epja/i2016-16158-2;%%
  %32 citations counted in INSPIRE as of 24 Dec 2019


%\cite{Brodsky:2000zc}
\bibitem{Brodsky:2000zc}
  S.~J.~Brodsky, E.~Chudakov, P.~Hoyer and J.~M.~Laget,
  %``Photoproduction of charm near threshold,''
  Phys.\ Lett.\ B {\bf 498}, 23 (2001)
  % doi:10.1016/S0370-2693(00)01373-3
  [hep-ph/0010343].
  %%CITATION = doi:10.1016/S0370-2693(00)01373-3;%%
  %65 citations counted in INSPIRE as of 24 Dec 2019


%\cite{Gryniuk:2016mpk}
\bibitem{Gryniuk:2016mpk}
  O.~Gryniuk and M.~Vanderhaeghen,
  %``Accessing the real part of the forward $J/\psi$-p scattering amplitude from $J/\psi$ photoproduction on protons around threshold,''
  Phys.\ Rev.\ D {\bf 94},  074001 (2016)
  % doi:10.1103/PhysRevD.94.074001
  [arXiv:1608.08205 [hep-ph]].
  %%CITATION = doi:10.1103/PhysRevD.94.074001;%%
  %12 citations counted in INSPIRE as of 24 Dec 2019


%\cite{Martynov:2001tn}
\bibitem{Martynov:2001tn}
  E.~Martynov, E.~Predazzi and A.~Prokudin,
  %``A Universal Regge pole model for all vector meson exclusive photoproduction by real and virtual photons,''
  Eur.\ Phys.\ J.\ C {\bf 26}, 271 (2002)
  % doi:10.1140/epjc/s2002-01058-5
  [hep-ph/0112242].
  %%CITATION = doi:10.1140/epjc/s2002-01058-5;%%
  %19 citations counted in INSPIRE as of 24 Dec 2019


%\cite{Martynov:2002ez}
\bibitem{Martynov:2002ez}
  E.~Martynov, E.~Predazzi and A.~Prokudin,
  %``Photoproduction of vector mesons in the soft dipole pomeron model,''
  Phys.\ Rev.\ D {\bf 67}, 074023 (2003)
  % doi:10.1103/PhysRevD.67.074023
  [hep-ph/0207272].
  %%CITATION = doi:10.1103/PhysRevD.67.074023;%%
  %23 citations counted in INSPIRE as of 24 Dec 2019


%\cite{Strakovsky:2019bev}
\bibitem{Strakovsky:2019bev}
  I.~Strakovsky, D.~Epifanov and L.~Pentchev,
  %``J/$\psi$p Scattering Length from GlueX Threshold Measurements,''
  arXiv:1911.12686 [hep-ph].
  %%CITATION = ARXIV:1911.12686;%%


%\cite{Kharzeev:1995ij}
\bibitem{Kharzeev:1995ij}
  D.~Kharzeev,
  %``Quarkonium interactions in QCD,''
  Proc.\ Int.\ Sch.\ Phys.\ Fermi {\bf 130}, 105 (1996)
  % doi:10.3254/978-1-61499-215-8-105
  [nucl-th/9601029].
  %%CITATION = doi:10.3254/978-1-61499-215-8-105;%%
  %38 citations counted in INSPIRE as of 24 Dec 2019


%\cite{Hatta:2018ina}
\bibitem{Hatta:2018ina}
  Y.~Hatta and D.-L.~Yang,
  %``Holographic $J/\psi$ production near threshold and the proton mass problem,''
  Phys.\ Rev.\ D {\bf 98},  074003 (2018)
  % doi:10.1103/PhysRevD.98.074003
  [arXiv:1808.02163 [hep-ph]].
  %%CITATION = doi:10.1103/PhysRevD.98.074003;%%
  %13 citations counted in INSPIRE as of 24 Dec 2019


%\cite{Fujii:1998tk}
\bibitem{Fujii:1998tk}
  H.~Fujii and D.~Kharzeev,
  %``Long range interactions of small color dipoles,''
  hep-ph/9807383.
  %%CITATION = HEP-PH/9807383;%%
  %8 citations counted in INSPIRE as of 24 Dec 2019


  %\cite{CAO:2020EicC}
\bibitem{CAO:2020EicC}
Xu Cao, Lei Chang, Ningbo Chang, {\it et al.}, {\it Electron Ion Collider in China (in Chinese)}, Nuclear Techniques, 2020, 43(2): 020001.
doi:10.11889/j.0253-3219.2020.hjs.43.020001
%X. Cao {\it et al.}, {\it Electron Ion Collider in China (EicC)}, to appear in Nuclear Science and Technology (2020).
%Xu Cao, Lei Chang, Ning-Bo Chang  {\it et al.} (in Chinese). Nuclear Science and Technology (in publication), 2020.

%\cite{Frankfurt:1998yf}
\bibitem{Frankfurt:1998yf}
  L.~L.~Frankfurt, M.~F.~McDermott and M.~Strikman,
  %``Diffractive photoproduction of $\upsilon$ at HERA,''
  JHEP {\bf 9902}, 002 (1999)
  % doi:10.1088/1126-6708/1999/02/002
  [hep-ph/9812316].
  %%CITATION = doi:10.1088/1126-6708/1999/02/002;%%
  %70 citations counted in INSPIRE as of 24 Dec 2019


%\cite{Chekanov:2002xi}
\bibitem{Chekanov:2002xi}
  S.~Chekanov {\it et al.} [ZEUS Collaboration],
  %``Exclusive photoproduction of J / psi mesons at HERA,''
  Eur.\ Phys.\ J.\ C {\bf 24}, 345 (2002)
  % doi:10.1007/s10052-002-0953-7
  [hep-ex/0201043].
  %%CITATION = doi:10.1007/s10052-002-0953-7;%%
  %295 citations counted in INSPIRE as of 24 Dec 2019


%\cite{Gittelman:1975ix}
\bibitem{Gittelman:1975ix}
  B.~Gittelman, K.~M.~Hanson, D.~Larson, E.~Loh, A.~Silverman and G.~Theodosiou,
  %``Photoproduction of the psi (3100) Meson at 11-GeV,''
  Phys.\ Rev.\ Lett.\  {\bf 35}, 1616 (1975).
  % doi:10.1103/PhysRevLett.35.1616
  %%CITATION = doi:10.1103/PhysRevLett.35.1616;%%
  %122 citations counted in INSPIRE as of 24 Dec 2019


%\cite{Laget:1994ba}
\bibitem{Laget:1994ba}
  J.~M.~Laget and R.~Mendez-Galain,
  %``Exclusive photoproduction and electroproduction of vector mesons at large momentum transfer,''
  Nucl.\ Phys.\ A {\bf 581}, 397 (1995).
  % doi:10.1016/0375-9474(94)00428-P
  %%CITATION = doi:10.1016/0375-9474(94)00428-P;%%
  %95 citations counted in INSPIRE as of 24 Dec 2019


%\cite{Sibirtsev:2004ca}
\bibitem{Sibirtsev:2004ca}
  A.~Sibirtsev, S.~Krewald and A.~W.~Thomas,
  %``Systematic analysis of charmonium photoproduction,''
  J.\ Phys.\ G {\bf 30}, 1427 (2004).
  % doi:10.1088/0954-3899/30/10/009
  %%CITATION = doi:10.1088/0954-3899/30/10/009;%%
  %11 citations counted in INSPIRE as of 24 Dec 2019


%\cite{Blin:2016dlf}
\bibitem{Blin:2016dlf}
  A.~N.~Hiller Blin, C.~Fern{\'a}ndez-Ram{\'i}rez, A.~Jackura, V.~Mathieu, V.~I.~Mokeev, A.~Pilloni and A.~P.~Szczepaniak,
  %``Studying the P$_c$(4450) resonance in J/$\psi$ photoproduction off protons,''
  Phys.\ Rev.\ D {\bf 94},  034002 (2016)
  % doi:10.1103/PhysRevD.94.034002
  [arXiv:1606.08912 [hep-ph]].
  %%CITATION = doi:10.1103/PhysRevD.94.034002;%%
  %47 citations counted in INSPIRE as of 24 Dec 2019


%\cite{Huang:2015uda}
\bibitem{Huang:2015uda}
  H.~Huang, C.~Deng, J.~Ping and F.~Wang,
  %``Possible pentaquarks with heavy quarks,''
  Eur.\ Phys.\ J.\ C {\bf 76},  624 (2016)
  % doi:10.1140/epjc/s10052-016-4476-z
  [arXiv:1510.04648 [hep-ph]].
  %%CITATION = doi:10.1140/epjc/s10052-016-4476-z;%%
  %65 citations counted in INSPIRE as of 24 Dec 2019


%\cite{Huang:2018wed}
\bibitem{Huang:2018wed}
  H.~Huang and J.~Ping,
  %``Investigating the hidden-charm and hidden-bottom pentaquark resonances in scattering process,''
  Phys.\ Rev.\ D {\bf 99},  014010 (2019)
  % doi:10.1103/PhysRevD.99.014010
  [arXiv:1811.04260 [hep-ph]].
  %%CITATION = doi:10.1103/PhysRevD.99.014010;%%
  %7 citations counted in INSPIRE as of 24 Dec 2019


%\cite{Lin:2018kcc}
\bibitem{Lin:2018kcc}
  Y.-H.~Lin, C.-W.~Shen and B.-S.~Zou,
  %``Decay behavior of the strange and beauty partners of $P_c$ hadronic molecules,''
  Nucl.\ Phys.\ A {\bf 980}, 21 (2018)
  % doi:10.1016/j.nuclphysa.2018.10.001
  [arXiv:1805.06843 [hep-ph]].
  %%CITATION = doi:10.1016/j.nuclphysa.2018.10.001;%%
  %8 citations counted in INSPIRE as of 24 Dec 2019


%\cite{Yang:2018oqd}
\bibitem{Yang:2018oqd}
  G.~Yang, J.~Ping and J.~Segovia,
  %``Hidden-bottom pentaquarks,''
  Phys.\ Rev.\ D {\bf 99},  014035 (2019)
  % doi:10.1103/PhysRevD.99.014035
  [arXiv:1809.06193 [hep-ph]].
  %%CITATION = doi:10.1103/PhysRevD.99.014035;%%
  %7 citations counted in INSPIRE as of 24 Dec 2019


%\cite{Shen:2017ayv}
\bibitem{Shen:2017ayv}
  C.-W.~Shen, D.~R{\"o}nchen, U.-G.~Mei{\ss}ner and B.-S.~Zou,
  %``Exploratory study of possible resonances in heavy meson - heavy baryon coupled-channel interactions,''
  Chin.\ Phys.\ C {\bf 42},  023106 (2018)
  % doi:10.1088/1674-1137/42/2/023106
  [arXiv:1710.03885 [hep-ph]].
  %%CITATION = doi:10.1088/1674-1137/42/2/023106;%%
  %12 citations counted in INSPIRE as of 24 Dec 2019


%\cite{Gutsche:2019mkg}
\bibitem{Gutsche:2019mkg}
  T.~Gutsche and V.~E.~Lyubovitskij,
  %``Structure and decays of hidden heavy pentaquarks,''
  Phys.\ Rev.\ D {\bf 100},  094031 (2019)
  % doi:10.1103/PhysRevD.100.094031
  [arXiv:1910.03984 [hep-ph]].
  %%CITATION = doi:10.1103/PhysRevD.100.094031;%%
  %4 citations counted in INSPIRE as of 24 Dec 2019

%\cite{Tanabashi:2018oca}
\bibitem{Tanabashi:2018oca}
  M.~Tanabashi {\it et al.} [Particle Data Group],
  %``Review of Particle Physics,''
  Phys.\ Rev.\ D {\bf 98}, 030001 (2018).
  % doi:10.1103/PhysRevD.98.030001
  %%CITATION = doi:10.1103/PhysRevD.98.030001;%%
  %3385 citations counted in INSPIRE as of 24 Dec 2019

%\cite{Wu:2019adv}
\bibitem{Wu:2019adv}
  J.-J.~Wu, T.-S.~H.~Lee and B.-S.~Zou,
  %``Nucleon resonances with hidden charm in ��p reactions,''
  Phys.\ Rev.\ C {\bf 100}, 035206 (2019)
  % doi:10.1103/PhysRevC.100.035206
  [arXiv:1906.05375 [nucl-th]].
  %%CITATION = doi:10.1103/PhysRevC.100.035206;%%
  %5 citations counted in INSPIRE as of 24 Dec 2019

  %\cite{Accardi:2012qut}
\bibitem{Accardi:2012qut}
  A.~Accardi {\it et al.},
  %``Electron Ion Collider: The Next QCD Frontier : Understanding the glue that binds us all,''
  Eur.\ Phys.\ J.\ A {\bf 52}, no. 9, 268 (2016)
  doi:10.1140/epja/i2016-16268-9
  [arXiv:1212.1701 [nucl-ex]].
  %%CITATION = doi:10.1140/epja/i2016-16268-9;%%
  %705 citations counted in INSPIRE as of 10 Mar 2020

  %\cite{Boer:2011fh}
\bibitem{Boer:2011fh}
  D.~Boer {\it et al.},
  %``Gluons and the quark sea at high energies: Distributions, polarization, tomography,''
  arXiv:1108.1713 [nucl-th].
  %%CITATION = ARXIV:1108.1713;%%
  %496 citations counted in INSPIRE as of 10 Mar 2020


%\cite{Xie:2020pc}
\bibitem{Xie:2020pc}
%\cite{Xie:2020niw}
%\bibitem{Xie:2020niw}
Y.~Xie, X.~Cao, Y.~Liang and X.~Chen,
%``Pentaquark $P_c$ electroproduction in $J/\psi +p$ channel in electron-proton collisions,''
[arXiv:2003.11729 [hep-ph]].
%0 citations counted in INSPIRE as of 14 Apr 2020


%\cite{Huang:2016tcr}
\bibitem{Huang:2016tcr}
  Y.~Huang, J.~J.~Xie, J.~He, X.~Chen and H.~F.~Zhang,
  %``Photoproduction of hidden-charm states in the $\gamma p \to \bar D^{*0} \Lambda^+_c$ reaction near threshold,''
  Chin.\ Phys.\ C {\bf 40}, no. 12, 124104 (2016)
%  doi:10.1088/1674-1137/40/12/124104
%  [arXiv:1604.05969 [nucl-th]].
  %%CITATION = doi:10.1088/1674-1137/40/12/124104;%%
  %17 citations counted in INSPIRE as of 10 Mar 2020

%\cite{Luders:2001nb}
\bibitem{Luders:2001nb}
  S.~L{\"u}ders,
  {\it A Measurement of the Beauty Production Cross Section via $B \rightarrow J/\psi X$ at HERA,}
  PhD thesis, ETH (2011).
  % doi:10.3929/ethz-a-004280689.
  %%CITATION = doi:10.3929/ethz-a-004280689;%%
  %1 citations counted in INSPIRE as of 24 Dec 2019


%\cite{Adloff:1999nr}
\bibitem{Adloff:1999nr}
  C.~Adloff {\it et al.} [H1 Collaboration],
  %``Measurement of open beauty production at HERA,''
  Phys.\ Lett.\ B {\bf 467}, 156 (1999)
  Erratum: [Phys.\ Lett.\ B {\bf 518}, 331 (2001)]
  % doi:10.1016/S0370-2693(99)01099-0, 10.1016/S0370-2693(01)01035-8
  [hep-ex/9909029].
  %%CITATION = doi:10.1016/S0370-2693(99)01099-0, 10.1016/S0370-2693(01)01035-8;%%
  %156 citations counted in INSPIRE as of 24 Dec 2019


%\cite{Ellis:1988sb}
\bibitem{Ellis:1988sb}
  R.~K.~Ellis and P.~Nason,
  %``QCD Radiative Corrections to the Photoproduction of Heavy Quarks,''
  Nucl.\ Phys.\ B {\bf 312}, 551 (1989).
  % doi:10.1016/0550-3213(89)90571-3
  %%CITATION = doi:10.1016/0550-3213(89)90571-3;%%
  %280 citations counted in INSPIRE as of 24 Dec 2019


%\cite{Frixione:1994dv}
\bibitem{Frixione:1994dv}
  S.~Frixione, M.~L.~Mangano, P.~Nason and G.~Ridolfi,
  %``Total cross-sections for heavy flavor production at HERA,''
  Phys.\ Lett.\ B {\bf 348}, 633 (1995)
  % doi:10.1016/0370-2693(95)00163-F
  [hep-ph/9412348].
  %%CITATION = doi:10.1016/0370-2693(95)00163-F;%%
  %205 citations counted in INSPIRE as of 24 Dec 2019




\end{thebibliography}
\end{document}